\documentclass[12pt]{article}

\usepackage{newtxtext,newtxmath}

\usepackage{graphicx}

\usepackage[letterpaper,margin=1in]{geometry}

\renewenvironment{abstract}
	{\quotation}
	{\endquotation}

\date{}

\makeatletter
\renewcommand{\fnum@figure}{\textbf{Figure \thefigure}}
\renewcommand{\fnum@table}{\textbf{Table \thetable}}
\makeatother

\usepackage{scicite}

\usepackage{url}

\newcommand{\TitledBox}[2]{%
    \vspace{1em} 
    \noindent
    \fbox{%
        \parbox{\dimexpr\linewidth-2\fboxsep-2\fboxrule}{%
            \textbf{#1}\\[0.5em] 
            #2 
        }%
    }%
    \vspace{1em} 
}

\def\scititle{
    Emergent Social Conventions and Collective Bias in LLM Populations\footnote{Preprint version of: Science Advances 11 (20), eadu9368 (2025).}
}
\title{\bfseries \boldmath \scititle}

\author{
    Ariel~Flint~Ashery$^{1}$,
	Luca~Maria~Aiello$^{2,3}$,
	Andrea~Baronchelli$^{1,4\ast}$,\and
	\small$^{1}$Department of Mathematics, City St George’s, University of London, Northampton Square, London, EC1V 0HB, UK.\and
	\small$^{2}$Computer Science Department, IT University of Copenhagen, Rued Langgaards Vej 7, 2300, Copenhagen, Denmark.\and
    \small$^{3}$Pioneer Centre for AI, 3 Øster Voldgade, 1350, Copenhagen, Denmark.\and
    \small$^{4}$The Alan Turing Institute, British Library, 96 Euston Road, London, NW1 2DB, UK.\and
	\small$^\ast$Corresponding author. Email: a.baronchelli.work@gmail.com
}

\begin{document} 

\maketitle

\vspace{0.2cm}
\begin{abstract} \bfseries \boldmath
Social conventions are the backbone of social coordination, shaping how individuals form a group. As growing populations of AI agents communicate through natural language, a fundamental question is whether they can bootstrap the foundations of a society.
Here, we present experimental results that demonstrate the spontaneous emergence of universally adopted social conventions in decentralized populations of Large Language Model (LLM) agents. We then show how strong collective biases can emerge during this process, even when agents exhibit no bias individually. Finally, we examine how committed minority groups of adversarial LLM agents can drive social change by imposing alternative social conventions on the larger population.
Our results show that AI systems can autonomously develop social conventions without explicit programming and have implications for designing AI systems that align, and remain aligned, with human values and societal goals.
\end{abstract}

\newpage
\section*{Introduction}
\noindent
Social conventions shape social and economic life, determining how individuals behave and their expectations~\cite{young1993evolution,ehrlich2005evolution,bicchieri2005grammar,gelfand2024norm}. They can be defined as unwritten, arbitrary patterns of behavior that are collectively shared by a group. Examples range from conventional greetings like handshakes or bows, to language and moral judgments~\cite{lewis1969convention,baronchelli2018emergence}. Recent numerical~\cite{steels1995self,baronchelli2006sharp} and experimental~\cite{centola2015spontaneous} results have confirmed the hypothesis that conventions can arise spontaneously, without the intervention of any centralized institution~\cite{lewis1969convention,bicchieri2005grammar,hayek1960constitution,sugden1989spontaneous}. Individual efforts to coordinate locally with one another can generate universally accepted conventions.

Do universal conventions also spontaneously emerge in populations of Large Language Models (LLMs), i.e., in groups of N simulated agents instantiated from an LLM? This question is critical for predicting and managing AI behavior in real-world applications, given the proliferation of LLMs using natural language to interact with one another and with humans~\cite{werfel2014designing, brinkmann2023machine, wu2023autogen}. Answering it is also a pre-requisite to ensure that AI systems behave in ways aligned with human values and societal goals~\cite{dafoe2021cooperative}.

A second key question concerns how the biases of individual LLMs influence the emergence of universal conventions, where ``bias" refers to an initial statistical preference for one option over an equivalent alternative in norm formation (e.g., individuals systematically preferring one name over another in a process leading to the population settling on a single name). Since collective processes can, in general, both suppress and amplify individual traits~\cite{mezard1987spin,castellano2009statistical}, answering this question is also relevant for practical applications. While most research has focused on investigating and addressing bias in one-to-one interactions between humans and LLMs~\cite{roselli2019managing,ferrara2023should,hu2023generative}, less attention has been given to how these biases evolve through repeated communications in populations of LLM agents and, ultimately, in mixed human-LLM ecosystems~\cite{dafoe2021cooperative}, even though the safety of a single LLM does not necessarily imply the safety of a multi-agent system~\cite{anwar2024foundational}.

Finally, a third question concerns the robustness of social conventions. Recent theoretical~\cite{xie2011social} and empirical~\cite{centola2018experimental} results have shown how a minority of adversarial agents can exert an outsized influence on the group, provided they reach a threshold or `critical mass'~\cite{kuran1998ethnic,kanter1977some,dahlerup2006story}. Investigating how conventions change through critical mass dynamics in a population of LLMs will help anticipate and potentially steer the development of beneficial norms in AI systems, while mitigating risks of harmful norms~\cite{baronchelli2024shaping}. It will also provide valuable models for how AI systems might play a role in shaping new societal norms to address global challenges such as antibiotic resistance~\cite{nyborg2016social} and the post-carbon transition~\cite{farmer2019sensitive}.

In this paper, we address these three key questions---on the spontaneous emergence of conventions, the role of individual biases, and critical mass dynamics---in populations of LLM agents. Drawing from recent laboratory experiments with human subjects~\cite{garrod1994conversation, centola2015spontaneous, centola2018experimental}, we follow the well-established practice of using coordination on a naming convention as a general model for conventional behavior~\cite{lewis1969convention, hume2000treatise, wittgenstein1958philosophical, garrod1994conversation, steels1995self, lazaridou2016multi}. In this setting, agents are endowed with purely local incentives and conventions may (or may not) emerge as an unintended consequence of individuals attempting to coordinate locally with one another. This sets our paper apart from the growing body of literature on LLM multi-agent systems, which has made considerable progress in complex problem-solving and world simulation but has primarily focused on goal-oriented simulations where LLMs either accomplish predefined group-level tasks or approximate human behavior in structured settings~\cite{guo2024large, ren2024emergence, horiguchi_evolution_2024, dafoe2021cooperative}. Unlike studies that use LLMs to predict human responses in social science experiments~\cite{hewitt2024predicting} or to simulate human societies~\cite{park2023generative,chuang2023simulating,han2024static}, our work does not treat LLMs as proxies for human participants but rather investigates how conventions emerge organically within a population of communicating AI agents as a result of their interactions~\cite{baronchelli2018emergence}. The emergence of conventions is a foundational element to any type of LLM multi-agent system~\cite{wu2023autogen,tran2025multi}, including but not limited to `in-silico' experiments to emulate human social networks~\cite{wang2024survey}. Here, we adopt a complex systems perspective~\cite{tsvetkova2024new}, rather than high-fidelity simulations of human interactions~\cite{rossetti2024social}, thereby minimizing the complexity of the experimental design to enhance the transparency of the result interpretation. Overall, our approach addresses recent calls for AI researchers to investigate how LLM agents may develop shared solutions to poorly defined social problems---such as creating language, norms, and institutions---to gain insights into the formation and stability of genuine cooperative AI systems~\cite{dafoe2021cooperative}.

\subsection*{Experimental Setting}
\subsubsection*{\textit{Background and Framework}}
 Our approach builds on Wittgenstein's general model of linguistic conventions, where repeated interactions lead to collective agreement between two players \cite{wittgenstein1958philosophical}. Theoretical extensions of this approach have argued that purely local interactions taking place on social networks can lead to population-wide, or `global', coordinated behavior \cite{young1993evolution,ehrlich2005evolution,skyrms2014evolution,baronchelli2018emergence}. Predictions for our study are based on the \emph{naming game model} of convention formation, where agents, aiming to coordinate in pairwise interactions, accumulate a memory of past plays, which they then use to ``guess" the words their subsequent partners will use~\cite{steels1995self,baronchelli2006sharp}. Extensive numerical and analytical work has shown how the model captures the rapid growth of universally shared social conventions in different settings~\cite{baronchelli2018emergence}.
Derived laboratory experiments involving human participants in naming games have provided the first empirical evidence for the spontaneous emergence of shared linguistic conventions~\cite{centola2015spontaneous}. A similar approach has confirmed these predictions by adopting more realistic input data within an application-driven setting \cite{lazaridou2016multi,michelrevisiting2023}.

The naming game framework has also been applied to study norm change and critical mass theory, which posits that committed minorities can overturn stable social conventions once their size reaches a tipping point, or `critical mass'. Theoretical models suggest critical masses between 10\% and 40\% of the population~\cite{xie2011social,niu2017impact}. Empirical evidence from controlled social coordination experiments, which closely follow the scheme described above, supports a 25\% threshold~\cite{centola2018experimental}. However, real-world observations reveal a wider range, with some studies proposing 30-40\% for gender conventions in corporate leadership~\cite{kanter1977some,grey2006numbers}, and others indicating that minorities as small as 0.3\% can trigger substantial linguistic and social changes~\cite{diani1992concept,gladwell2010small,amato2018dynamics,farmer2019sensitive}. 

\subsubsection*{\textit{Experimental Setup}} 
A simulation trial consists of a population of $N$ interacting agents. At each time-step, two agents are randomly selected for interaction. Interaction rules are specified by prompting the LLM agent (see next section). From a multi-agent perspective, each agent outputs a convention, or `name', from a pool of finite size $W$, and these outputs are compared to determine coordination. The prompt specifies that if the conventions match, the game score of the agent is incremented, and if they do not match, it is decremented. In either scenario, the game scores of both agents change by the same amount. This implements an incentive for coordination in pairwise interactions, while no incentive promotes global consensus. Moreover, the prompt does not specify that agents are part of a population or provide any detail on how the interaction partner is selected from a group. Importantly, the prompt provides the LLM agent with a `memory' storing details about the past $H$ interactions they participated in, including their co-player's convention choice, their own convention choice, whether the interaction was successful or not, and their own accumulated score over these $H$ interactions. The memory is initialized as empty so that, in the first interaction, the output is a random convention chosen from the pool of available names.

Finally, in the experiments on norm change and critical mass theory, we introduce a small number of adversarial agents (i.e., a `committed minority') into each population, who consistently promote a novel alternative at every interaction and irrespectively of their history~\cite{xie2011social,centola2018experimental}. These dynamics reflect common types of online interactions where community members engage directly with a large, often anonymous population---using chat interfaces or messaging technologies---leading to the adoption of linguistic and behavioral conventions that enable effective coordination with other participants' expectations~\cite{centola2015spontaneous,backstrom2006group,kooti2012emergence,centola2018experimental}. Here, we simulate these social interactions with four different LLM models: \textbf{Llama-2-70b-Chat}, \textbf{Llama-3-70B-Instruct}, \textbf{Llama-3.1-70B-Instruct}, and \textbf{Claude-3.5-Sonnet} (see \textit{Materials and Methods}).

\subsubsection*{\textit{Prompting}}
Interactions within the game take place in the form of a series of text-based moves. In each interaction, the LLM agent is given a text prompt comprised of a system prompt and a user input prompt. The system prompt contains all information about the game. The user input requests the agent to predict a player's next action based on the history of choices in the $H$ most recent interactions. This positions the agent as an external observer of the game, tasked with forecasting the upcoming round. In practice, these decisions dictate the state of play. Agents do not receive information about the players' identities or personalities, such as whether they are rational actors. Consequently, we can interpret the agent's recommendations as their de-facto participation in the game.

The system prompt (see \textit{Materials and Methods}) is designed such that the agent’s output follows a consistent format, from which we can extract its decision. Following previous works on LLMs' cognitive abilities~\cite{kojima2022large}, we prompt the agent to `think step by step' and to explicitly consider the history of play. The prompt thus encourages agents to make a decision based on their previous experience, but provides no instruction as to how it should be used in the decision making process. Agents are asked to select a name from the name pool, which is presented to them as a list of $W$ unique letters sampled from the English alphabet. Ordering bias is removed by randomizing the list of presented letters for each player at every interaction. A successful interaction garners equal rewards for the participating agents, whereas a failure to coordinate results in a penalty. In the absence of human guidance, LLMs are notoriously bad at arithmetic~\cite{huang2023large}. To avoid decision errors based on a misjudgment of the game state, we explicitly provide the agent with both the payoff they obtained at each round and their cumulative score within memory range. Lastly, to ensure that the responses generated by the LLM are correctly guided by the prompt and not merely the result of random hallucinations~\cite{xu2024hallucination}, we have implemented a meta-prompting strategy to assess the LLM's understanding of the given instructions. This practice, previously used in evaluating LLMs within game-theoretical frameworks~\cite{fontana2024nicer}, consists of posing a series of text comprehension queries to the LLM and evaluating the precision of its responses. The LLMs subjected to our testing demonstrated good comprehension capabilities (see Fig.~\ref{fig: meta prompting}).

\subsection*{Results}
To balance experimental time, which should allow for multiple repetitions, with parameters that provide agents a rich set of alternatives and meaningful awareness of their history, we set the name pool size to $W=10$ and the individual memory length to $H=5$ for populations of $N=24$ agents, unless otherwise specified. The results presented below remain robust with respect to variations in these parameters (see Fig.~\ref{fig: conv robustness}).

\subsubsection*{\textit{Spontaneous Emergence}}
\begin{figure}[!h]
\centering
        \includegraphics[width=0.6\linewidth]{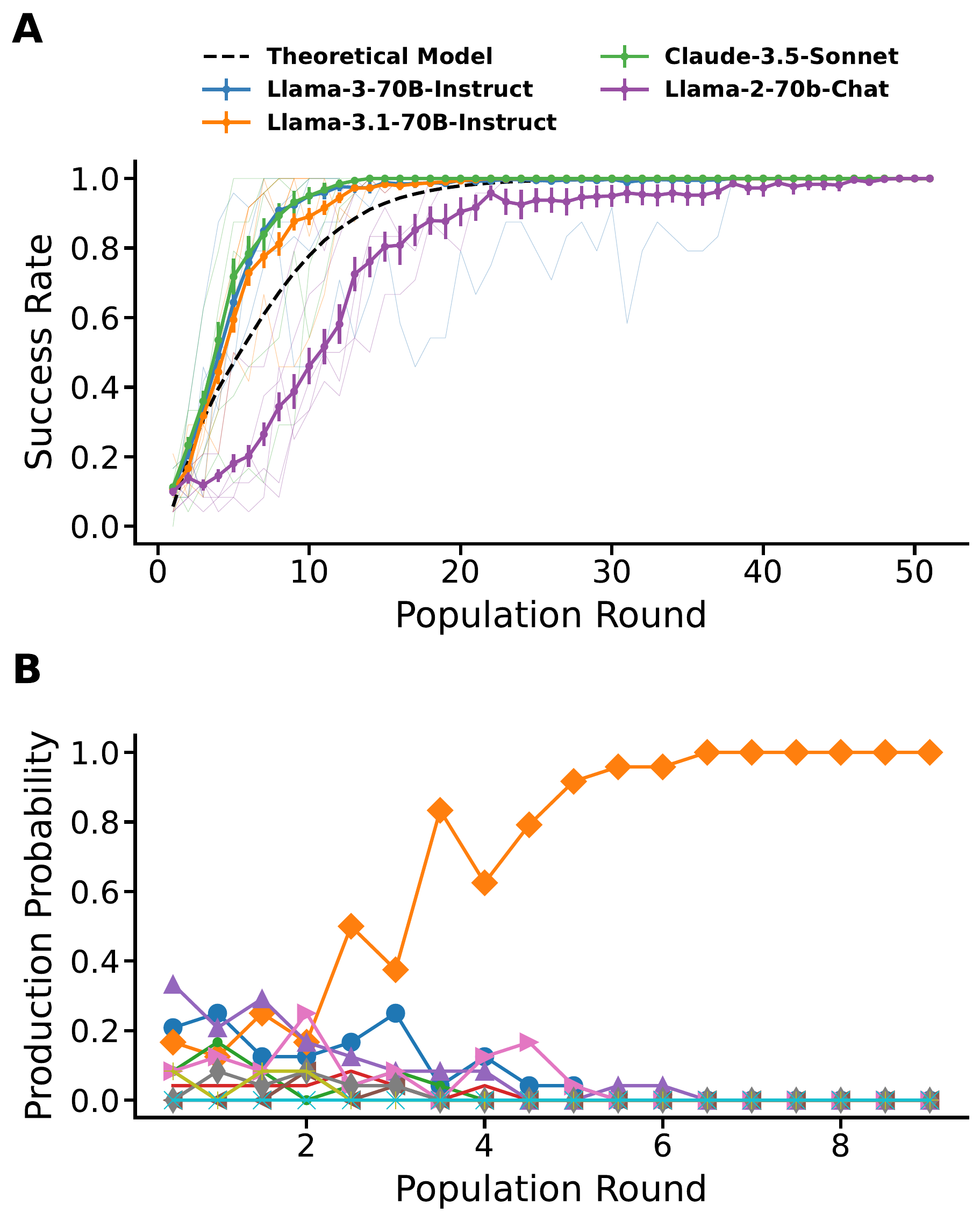}
        \caption{\textbf{The spontaneous emergence of  conventions}. \textbf{(A)} The success rate---i.e., the probability of observing a success at a given time---for population size $N=24$ and a name pool of size $W=10$, for each of the four models. Thick lines represent average curves obtained from 40 experimental runs, while thin lines are representative individual runs. To improve visibility, we only show 5 individual trajectories for each LLM. The black, dashed line shows the success rate of the theoretical minimal naming game model, averaged over 10,000 runs under the same constraints. 
        \textbf{(B)} Word competition in a single run in a population of \textbf{Llama-3.1-70B-Instruct} agents. Different markers and colors represent the trajectories of unique conventions. Each data point is a bin averaging the past interactions up until the preceding bin boundary. Error bars indicate standard error of the mean.
        }
    \label{fig:emergence}
    
\end{figure}

Fig.~\ref{fig:emergence} shows that group-wide linguistic conventions spontaneously emerge across all models. The dashed black line shows that the theoretical model (see Supplementary Text for a description) captures the dynamics generated by the LLM populations.

Initial steps have a low probability of success because the random pairing of agents makes repeated interactions improbable, thus preventing the formation of `neighborhoods' of entrenched behavior. However, these local dynamics lead to a disorder-to-order transition toward a consensus state where every agent systematically outputs the same name, i.e., where a global convention has emerged. The fact that the population converges to one of many possible alternatives characterizes the transition as a case of symmetry breaking \cite{baronchelli2006sharp}. This interpretation is further supported by examining the space of competing alternatives, shown in Fig.~\ref{fig:emergence}B. After an initial period in which several names are nearly equally popular, a single convention rapidly becomes dominant, transitioning the system into a ``winner-take-all" regime. Interestingly, the speed of convergence is similar across models: a shared social convention is established by population round 15 in all cases, except for \textbf{Llama-2-70b-Chat}, the least advanced LLM considered.

A natural question is whether consensus on a global convention also occurs (i) for larger population sizes, where the probability of repeated interactions is reduced, and (ii) when the number of competing alternative conventions increases, which could potentially complicate even local convergence. Fig.~\ref{fig: conv robustness} shows that populations as large as $N=200$ agents reach consensus and that a shared convention emerges for a name pool as large as $W=26$, demonstrating the robustness of the convergence process. Remarkably, larger populations reach consensus at a comparable speed, measured in terms of population rounds, while the effect of the name pool size $W$ is more nuanced, though not dramatic. In the next section, we examine how the composition of the available pool of conventions affects convergence.

\subsubsection*{\textit{Collective bias in convention selection}}
\begin{figure}
\centering
\includegraphics[width=0.6\linewidth]{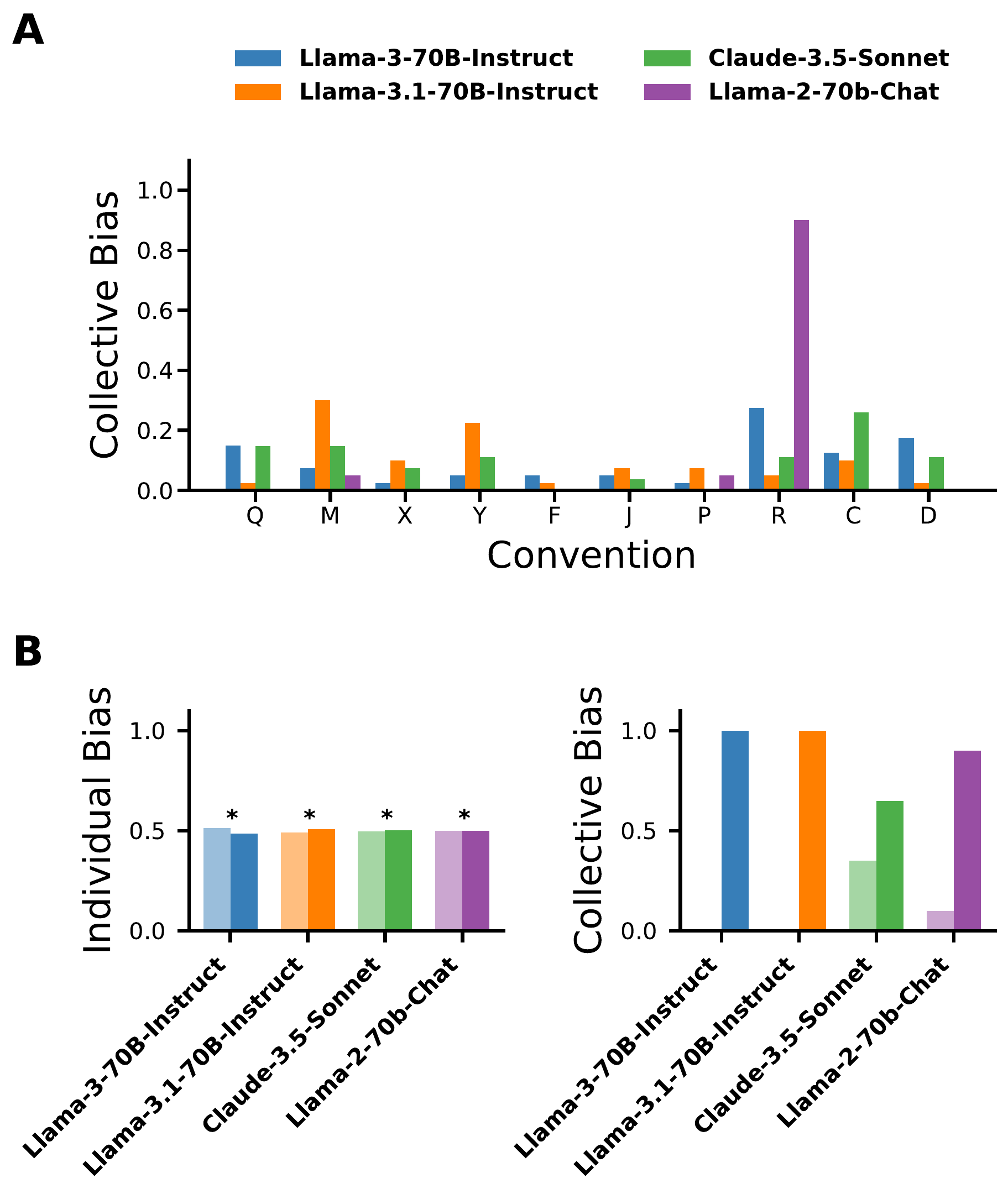}
\caption{\textbf{Collective Bias in Convention Selection}. \textbf{(A)} Distribution of consensus conventions, for a name pool of size $W=10$ ($N=24$). 
Results of 40 runs for the \textbf{Llama-3-70B-Instruct} and \textbf{Llama-3.1-70B-Instruct} models, and 27 and 20 runs for \textbf{Claude-3.5-Sonnet} and \textbf{Llama-2-70b-Chat}, respectively. The collective dynamics systematically amplify individual biases (shown in Fig.~\ref{fig: ten individual bias}). \textbf{(B)}, Individual vs Collective bias for $W=2$, name pool $\{Q, M\}$. Left panel:  probability of selecting either convention for  agents with no prior memory ($Q$: lighter hue, $M$: darker hue). Raw values reported in Table~\ref{tab:2A data}. Asterisks (*) indicates that there is insufficient evidence to reject the null hypothesis that the model is unbiased at the 5\% significance level (calculated using an exact Binomial test from 10,000 samples per model, apart from \textbf{Llama-3-70B-Instruct} which had 5,000 samples, see \textit{Materials and Methods}). Corresponding p-values for the models (from left to right) are $p$ = 0.068, 0.116, 0.757, and 0.849. 
Right: the proportion of runs ($40$) that resulted in consensus on the respective convention. Raw values reported in Table~\ref{tab:2B data}.
}
\label{fig:conv counts}
\end{figure}

Having established that social conventions emerge, a natural question arises: what are these conventions? The single Latin alphabet letters available in the name pool are all equally valid as global conventions, and so we would expect them to all to have the same probability to become the accepted social convention, as supported by the theoretical model~\cite{baronchelli2006sharp}  (see also Supplementary Text). However, the experimental results present a different picture (Fig.~\ref{fig:conv counts}A). The probability that a particular name becomes the social convention is neither uniform nor deterministic. Some names appear to have a pronouncedly higher likelihood of becoming the adopted convention than others. This pattern holds across models, although the preferred names vary between models.

Two hypotheses could explain the observed behavior. The selection process may be non-uniform due to \emph{(i)} intrinsic model (i.e., individual, single-agent) biases or \emph{(ii)} prompt features, specifically the order in which names in the name pool are presented to the agents, as noted in a different context~\cite{demarzo2023emergence}. The latter hypothesis can be discarded since, as mentioned above, the names are presented to the agents in a list in randomized order for each agent and at every interaction.

Having ruled out the order of name presentation as a factor, we can focus on the role of individual (i.e., single-agent) biases in shaping collective behavior. 
The hypothesis that \emph{individual} bias can be responsible for a \emph{collective bias} is supported by the theoretical model. When the theoretical model is run with only two names, a bias towards a particular name quickly results in unilateral convergence on that name at the population level (see Fig.~\ref{fig: theoretical production}). The speed of convergence depends on the size of the bias.

To test this intuition in our experiment, we examine the selection preferences of individual agents during their first round, when they have no prior memory. We find that individual biases are indeed possible. For example, when agents can choose any letter from the complete English alphabet, the population systematically converges on the letter `A' because individual agents overwhelmingly prefer to select it over all other letters, even without prior memory (see Fig.~\ref{fig: latin emergence}).
However, a similar test on the frequency of name selection by agents with no prior memory for the case of Fig.~\ref{fig:emergence},
where the name pool contains ten elements but not the letter `A', yields mixed results. Under these conditions, individual \textbf{Llama-2-70b-Chat} and \textbf{Claude-3.5-Sonnet} agents are unbiased across conventions in this name pool ($\chi^{2}$-test, $p$ = 0.100, 0.410), whereas individual \textbf{Llama-3/3.1-70B-Instruct} agents exhibit a significant statistical skew in their name selections (see Fig.~\ref{fig: ten individual bias}). In all cases, the final consensus distribution shows that specific names are favored as a consensus option, even if they appeared to be less likely to be selected in the initial step (Fig.~\ref{fig:conv counts}A). Thus, both social conventions and collective biases in the selection process emerge also in absence of individual biases.

The findings suggest that collective bias may stem from the convention formation process itself, as agents are exposed to diverse memory states with different name combinations and success-failure sequences. To test this hypothesis, we focus on the case of a name pool size $W=2$, since tracking potential confounders of bias becomes impractical as the space of possible names increases. Fig.~\ref{fig:conv counts}B shows that across all models, although agents are initially unbiased, local communication and coordination lead to a collective bias toward a specific convention, which we term the `strong convention' (as opposed to its `weak' counterpart). This finding is consistent across various convention combinations (see Fig.~\ref{fig: binary emergence}).

We examine the microscopic contributions to collective bias in Table~\ref{table: microdynamics}. The top row of Table~\ref{table: microdynamics} shows a case where there is no individual bias towards a particular name in the first interaction ($p$ = 0.11 $>$ 0.05, indicating that the evidence is not strong enough to reject the hypothesis that the agent is unbiased). In the second interaction, agents have some memory influencing their decision, but the observed outcome probability remains symmetric ($p$ = 0.110). We observe that if an agent succeeds in the first interaction, it will almost surely continue to use the successful name in the next interaction ($99.4\%$ of the time in the data in Table~\ref{table: microdynamics}, with similar results in real simulations and for other models). However, if an agent fails, it will almost surely switch names ($97.3\%$ of the time). 
In all tested cases with $W=2$, and across all models, an asymmetric selection bias emerges by the agent’s third interaction, distinguishing between the ‘weak’ and ‘strong’ conventions. For the model and name pool reported in Table~\ref{table: microdynamics}, agents at this stage are more likely to choose the strong name in 5 out of the 8 most expected memory states. Crucially, the agent's strategies are not symmetric under a re-labelling of the conventions in the memory state. The most egregious example of this from Table~\ref{table: microdynamics} is $p(M | \{1: M, Q; 2: Q, M \})$ and $p(Q | \{1: Q, M; 2: M, Q \})$, which are equal to $0.848$ and $0.451$, respectively. In subsequent interactions, agents are more likely to encounter the strong name in successful interactions, reinforcing its use and ultimately leading to consensus on that name as the social convention.

In summary, our results suggest that a collective bias may also emerge also from repeated interactions among agents who, when tested in isolation (i.e., in interaction 1), appear to be unbiased in their decision making.
It is important to emphasize that this dynamically emerging bias is not required for the spontaneous emergence of a convention. The collective and individual biases of these agents drive the consensus towards \textit{particular} conventions. For reference, the theoretical model produces conventions without any individual bias, but accommodates it at the individual level to explain the dominance of specific conventions over competing alternatives \cite{baronchelli2006sharp,baronchelli2018emergence} (see Fig.~\ref{fig: theoretical production}). 
In LLMs, on the contrary, we observe that bias emerges when agents develop diverse memory states, which form through a collective process of agent-to-agent communication. Our results are robust with respect to variations in the prompt and convention labels and hold also in non-fine-tuned LLMs (see Supplementary Text).

\subsubsection*{\textit{Tipping Points and Critical Mass}}

Social conventions are steady states of the system: once a global convention spontaneously emerges, the population adheres to it indefinitely (see Fig.~\ref{fig: stability robustness}).
A natural question concerns the stability of such steady states: how resistant is a convention to deliberate efforts to overturn it?
To address this question, we investigate whether a committed minority can `flip' an equilibrium consensus on a convention. We consider the scenario in which a population has long converged on a convention and every agent has solely observed that convention in the past $H$ interactions (which were, therefore, all successful). We then introduce a `committed minority' of agents producing an alternative convention \cite{xie2011social, centola2018experimental}. These committed agents follow a fixed strategy and use the alternative convention at all times. We test populations using the same two-name ($W=2$) conditions as in our convergence experiments. We simulate a consensus on each name per combination and introduce its complementary name as an adversary.

\begin{figure}
\centering
\includegraphics[width=0.6\linewidth]{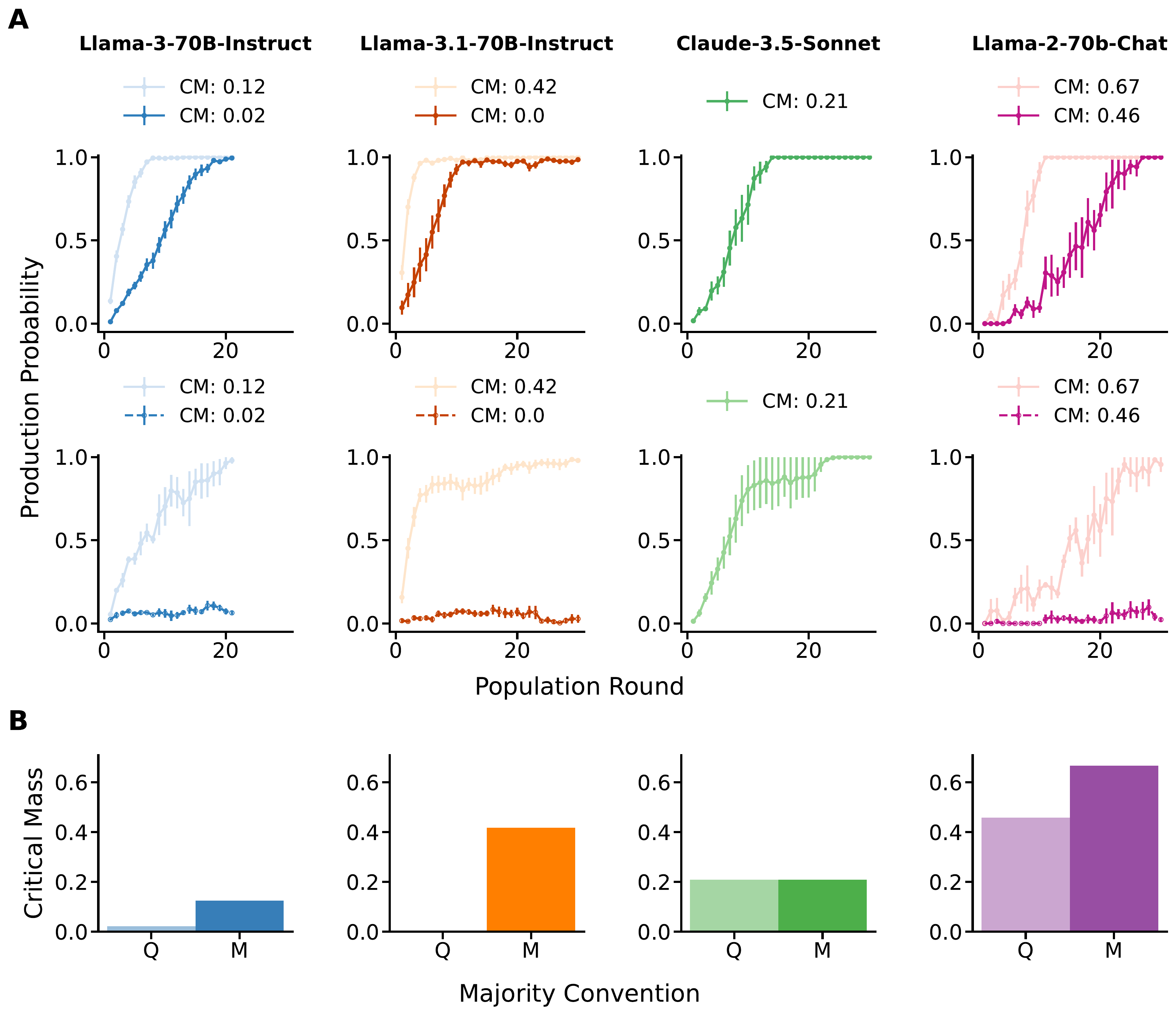}
\caption{\textbf{Committed minority and critical mass dynamics.} Populations of $N = 24$ agents ($N = 48$ for \textbf{Llama-3-70B-Instruct}) were initialized in two conditions, with complete consensus on either the weak ($Q$) or strong ($M$) convention ($W = 2$). Each agent's memory exclusively stored one convention in each setting, with memory length $H = 5$ ($H = 3$ for \textbf{Llama-3-70B-Instruct}). \textbf{(A)} The average probability of producing the alternative convention when the majority holds the weak (top) or strong (bottom) convention. The legend shows the size of the committed minority (CM). Bold (faint) lines represent the production probability when the CM reaches the critical mass needed to flip the majority on the strong (weak) convention. Solid lines with filled circles indicate that all trials achieved population consensus on the alternative convention (95\% success rate in the past $3N$ rounds).
\textbf{(B)} Critical mass needed to flip the majority for each model. Raw values reported in Table~\ref{tab:3B data}. Error bars indicate standard error of the mean.}

\label{fig: tipping points}
\end{figure}

In Fig.~\ref{fig: tipping points}, we show that when the committed minority reaches the critical threshold, the whole population adopts their convention. Below this threshold, the population settles into a mixed state, as committed agents always use the minority convention. Interestingly, the critical mass of the committed minority needed to trigger a new convention depends on the convention itself. 
The stronger name (i.e., the name more likely to become the social convention had we started with no prior memory, as seen in the previous section) requires a larger committed minority to be overturned. Conversely, a smaller number of adversarial agents can overturn a consensus on the weaker name.

Interestingly, the relative strength of the two conventions can vary so widely depending on the LLM that committed groups as small as 2\% (\textbf{Llama-3-70B-Instruct}) or as large as 67\% (\textbf{Llama-2-70b-Chat})---effectively no longer a minority---were observed (see Fig.~\ref{fig: tipping points}). In \textbf{Llama-3.1-70B-Instruct} populations, the bias is so strongly weighted against the weaker convention that the population spontaneously switches to the alternative, stronger convention without requiring any committed agents at all. Relative strength can be understood by considering the limits of an agent's exploration, i.e., the likelihood that their output deviates from the strong (weak) social convention as the game unfolds (see Supplementary Text). As the population converges toward the strong convention, agents quickly reach memory configurations that resist further exploration, making the consensus steady state robust. In contrast, weaker conventions coexist with a greater propensity for exploration among agents. Similar dynamics take place when the system is perturbed from a consensus steady state on the strong convention. Adopting a dynamical systems perspective, we can say that the basin of attraction of the strong convention is both larger and deeper than that of the weaker convention, as it attracts more system configurations and makes it more difficult for the system to escape (see Supplementary Text).

\subsection*{Discussion}
Our findings show that social conventions can spontaneously emerge in populations of Large Language Models (LLMs) through purely local interactions, without any central coordination. These results reveal how the process of social coordination can give rise to collective biases, increasing the likelihood of specific social conventions developing over others. Importantly, this collective bias is not easily deducible from analyzing isolated agents, and its nature varies depending on the LLM model used.
Additionally, our work uncovers the existence of tipping points in social conventions, where a minority of committed agents can impose their preferred convention on a majority settled on a different one. The critical size of this committed minority is influenced by at least two factors: the interplay between the majority's established convention and the minority's promoted alternative, and the specific LLM model employed. 

Within the expanding field of LLM multi-agent systems~\cite{guo2024large}, 
multi-agent experiments with AI agents simulating opinion dynamics models suggest that LLMs are able to reach consensus in groups without any incentive, although this is limited by group size~\cite{de_marzo_language_2024}. In this context, our study presents a flexible benchmarking framework to detect the hidden higher-order biases that could arise from complex interactions in social LLM experiments~\cite{bail_can_2024}. Our results on norm change could stimulate research into similar dynamics within the framework of cultural evolution, particularly in chains of communicating agents~\cite{perez2024llms}. Game theoretical approaches would naturally allow investigation of asymmetric payoffs' effects on collective consensus, potentially contrasting individual biases with explicit collective goals~\cite{kearns2009behavioral, duan2024gtbench,davidson2024evaluating}.
Further promising research avenues include developing frameworks to promote the emergence of specific conventions~\cite{ren2024emergence} and higher-order social norms~\cite{lazaridou2018emergence, horiguchi_evolution_2024}, as well as testing interactions between agents based on different LLMs within populations.

It is important to delimit the scope of our findings while highlighting possible avenues for future work.
First, our results reveal key aspects of norm dynamics in populations of LLMs within an experimental setup that is, unavoidably in LLM research, reliant on several parameters including the LLM model, the prompt, and specific conventions. While rigorous testing, including meta-prompting and experiment repetitions using different parameters, confirms the robustness of the results in this context, an important aspect of future work will consist of generalizing the results to different controlled experimental settings. In this context, scaling to larger populations and semantic spaces should also be investigated~\cite{michelrevisiting2023,de_marzo_language_2024}.
Second, we considered only unstructured populations where interacting pairs are randomly selected. A straightforward yet crucial extension of this work consists of embedding the population in more realistic social networks, which may have a profound impact on the collective dynamics~\cite{baronchelli2018emergence,han2024static}, as well as considering microscopic interactions involving more than two agents~\cite{iacopini2022group}.
Third, to bridge the gap between synthetic experiments and real-world applications, an exciting frontier for future study lies in considering more realistic conventions---such as moving from alphabet letters to sensitive human norms related to gender, race, and other social categories. Last, simulated cooperative games played by AI agents may also prove useful for tuning the agentic behavior towards desirable outcomes. This could be potentially achieved through Multi-agent Reinforcement Learning~\cite{sun2024llm} or, in games for which a clear optimal strategy can be defined, by integrating strategic reasoning into the agent's decision workflow through external knowledge bases or Bayesian reasoning modules~\cite{hua2024game, gemp2024steering}.

An important point concerns the dialogue between our results on AI agents and the current understanding of social convention dynamics in humans. 
On the one hand, our results showed qualitative similarities between the collective dynamics of AI and human subjects, concerning both the emergence of shared norms and critical mass dynamics. On the other hand, we unveiled what appear to be LLM-specific phenomena regarding collective bias, affecting both the emergence and resilience of conventions,  which call for further human testing.
These indications are important because assessing the similarities and differences between artificial and human societies in such a foundational aspect as norm dynamics has implications for digital-twin synthetic modeling and applications~\cite{rossetti2024social}. For example, if AI agents behaved exactly like humans, then synthetic testing of norm dynamics under collective stresses---such as pandemics, terrorism, or wars---would be justified. If, on the other hand, LLM-specific dynamics proved to be substantial (e.g., if evidence of collective bias were further confirmed), then using these agents as simulations of human social systems or deploying these agents in social settings such as social media would require additional care. In particular, it is crucial to develop techniques to systematically identify discrepancies between LLM outputs and the expected human behavior~\cite{gligoric2024can}, to then correct them with statistical techniques~\cite{egami2023using} or by keeping human judgments in the loop~\cite{li2023coannotating}. Addressing these points is a key endeavor for the future, with far-reaching implications. Next steps involve further investigating convention dynamics in human and AI populations as well as in mixed LLM-human ecosystems, both in laboratory settings and eventually in natural environments like social media.

Our work also underscores the ethical challenges of bias propagation in LLMs. Despite their rapid adoption, these models pose serious risks, as the vast, unfiltered Internet data used to train them can reinforce and amplify harmful biases, disproportionately harming marginalized communities~\cite{gallegos_bias_2024}. Accordingly, a central focus of the alignment research community has been to improve LLM performance in individual bias tests~\cite{wang_aligning_2023, ouyang_training_2022}. However, our findings reveal that alignment must also be tested at the group level, where collective biases can emerge and persist.

Finally, understanding how AI systems spontaneously develop conventions and more sophisticated norms without explicit programming is a critical first step for predicting and managing ethical AI behavior in real-world applications while ensuring agent alignment with human values and societal goals.  It is also crucial for safeguarding AI agents from potential attacks.
In particular, tipping points in norm dynamics present both opportunities, such as addressing global challenges~\cite{nyborg2016social,farmer2019sensitive}, and risks, particularly if exploited for social control~\cite{king2017chinese}. Our findings highlight potential vulnerabilities in multiagent systems, which could be exploited through injection attacks to influence the emergence of specific norms~\cite{lee2024prompt}. Recognizing these risks, studying collective LLM behavior is crucial for assessing potential harms from the integration of AI agents into applications and for developing effective mitigation strategies. Moreover, efforts to measure and instill human social norms in LLMs have so far yielded mixed results~\cite{yuan_measuring_2024, hammerl_multilingual_2022}, and as of yet AI agents struggle to represent multiple cultures~\cite{ramezani_knowledge_2023} and continuously evolving social norms~\cite{li_agent_2024, shen_shaping_2023, baronchelli2024shaping}. We argue that the challenge extends beyond merely detecting `undesirable behavior', to understanding the evolution of social norms held by agents~\cite{baronchelli2024shaping}. In this light, our work represents a first step towards a better understanding of norm and bias dynamics in populations of LLMs, and we anticipate that it will be of interest to researchers and practitioners working to make AI a tool for societal good.

\newpage
\subsection*{Materials and Methods}
\subsubsection*{Prompt}
The system prompt comprises of three components: \emph{i)} a fixed prompt that outlines the game’s rules, including the payoff structure and the player’s objective, \emph{ii)} a dynamic memory prompt that provides contextual information about the state of play within the player's memory range, and \emph{iii)} an instructional prompt that provides information for how the agent should format its response. The user prompt asks the agent to select a name to use in the current interaction. We use zero-shot prompting to directly extract the agent’s name decision in response to the state of play. We do not provide instructions as to how agents should decide their next move, nor do we present them with example strategies. We ask the agent to behave in a self-interested manner, and the only part of the prompt in which we suggest to the agent that it should consider partaking in coordination is when we state that the agent’s objective is to ‘maximize their own accumulated point tally, conditional on the behavior of their co-player’. We apply fixed payoffs for successful and failed interactions, set at +100 and -50 points respectively.

\subsubsection*{\textit{Models and APIs}}

For our experiments, we use homogeneous populations of agents instantiated from the following LLMs: \textbf{Llama-3-70B-Instruct}, \textbf{Llama-3.1-70B-Instruct}, \textbf{Llama-2-70b-Chat} (in 4-bit quantization format), and \textbf{Claude-Sonnet-3.5} (see Table~\ref{tab:model_versions} for specific versions). All Llama family models are open-sourced LLMs, released under a commercial use license (\url{https://ai.meta.com/llama/license/}). We use versions of the Llama 3 family models hosted by Hugging Face, which we access through the Inference API (\url{https://huggingface.co/inference-api/serverless}). We quantize \textbf{Llama-2-70b-Chat} into a 4-bit version using Hugging Face's \textsc{Transformers} library (\url{https://huggingface.co/docs/transformers}), and run the model locally using a single A100 GPU. In auto-regressive LLMs, each newly generated word is produced based on previously inputted and generated words, and so the sequence of generation matters. More precisely, the probability distribution for predicting the next word is conditional on the product of all previous word probability distribution. To mimic LLMs deployed in real-world applications, we demand all agents in our experiments to behave non-deterministically by fixing them with a non-zero constant temperature. This means that for each agent the next generated word is randomly selected from the conditional probability distribution. We use $K$-sampling to restrict the probability distribution of the next word to the next $K$ most likely words, thus increasing the likelihood of high probability words and decreasing the likelihood of low probability words which are outside of the name pool (see Table~\ref{table:llm_parameters} for all parameter values).

\subsubsection*{\textit{Measuring Individual Bias}}
We quantify the individual bias of agents by measuring the number of times each convention was produced in the first round of the game, when their memory inventory is empty, over $T$ trials. Experiments with $W=2$ are effectively a Bernoulli trial, and so we measure whether the agent is biased by performing a two-tailed exact Binomial test with the observed proportions. We calculate the p-value using a null probability of $0.5$, and reject the hypothesis that the model is biased if $p<0.05$. For the case of $W=10$, we perform a $\chi^{2}$-test, and also test the null hypothesis that the model is neutral in its convention selection. Thus, we use the expected value $0.1T$ in our calculations, and again reject the null hypothesis that the model is unbiased if $p<0.05$.

\subsubsection*{\textit{Committed Minorities}}

To determine the critical size of the committed minority, we identify the point at which the majority consensus is overturned. A consensus flip occurs when 95\% of the past $3N$ interactions succeed after the introduction of the committed minority. For \textbf{Llama-3-70B-Instruct}, we tested the smallest minority needed to overturn a weak convention majority, then repeated the experiment with a strong convention majority to measure the critical mass within the same time frame. For other models, the critical mass threshold is defined as the minimum proportion of committed agents that is required to flip the consensus within 30 population rounds. These criteria account for potential fluctuations in non-deterministic agent decisions.

\newpage

\begin{table}
    \centering
    \includegraphics[width=0.6\linewidth]{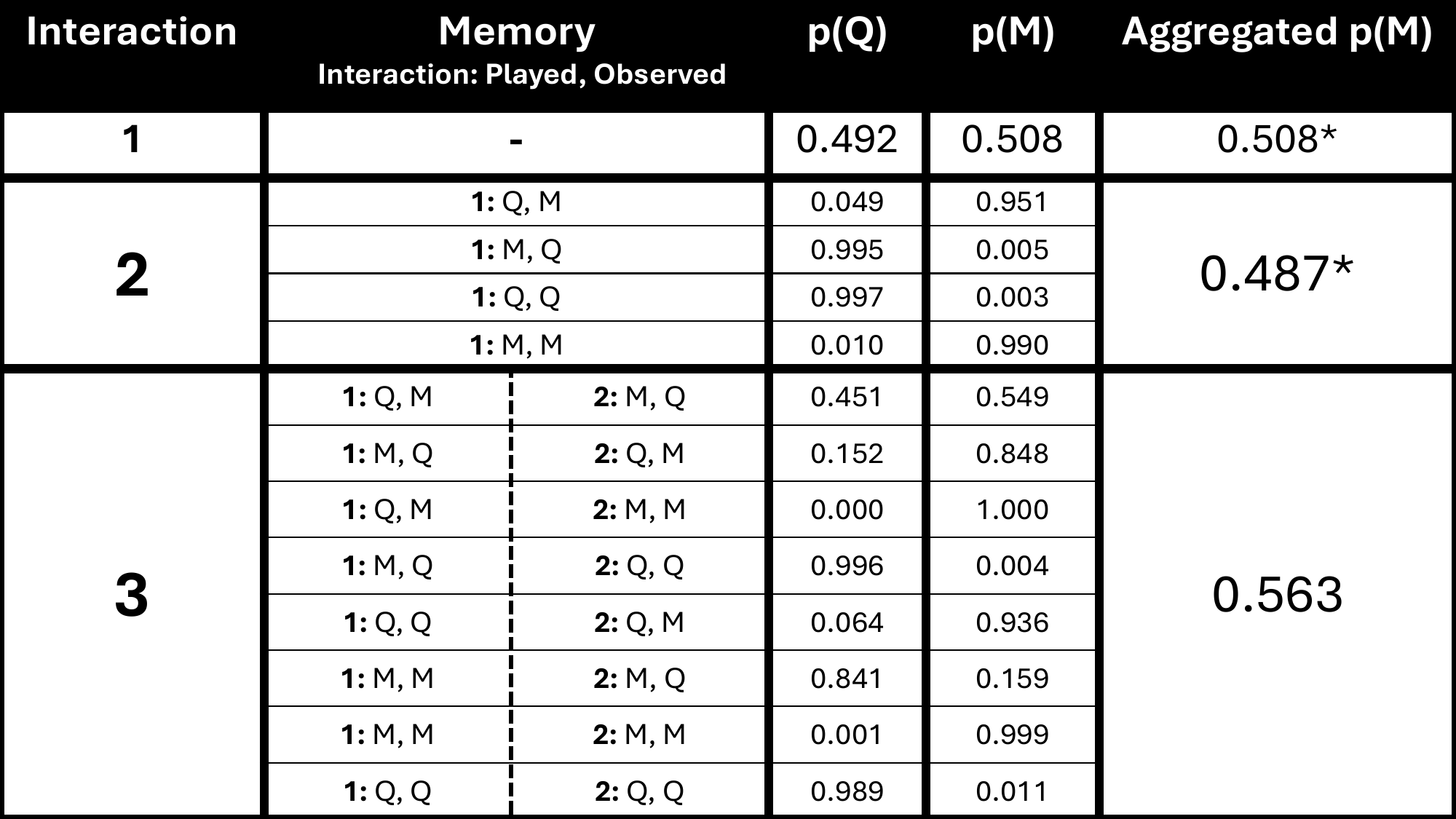}
    \vspace{5mm}
    \caption{\textbf{The origin of collective bias.} \textmd{The strategies of a \textbf{Llama-3.1-70B-Instruct} agent in the early phases of the experimental setting up to the third interaction, with $W = 2$ and a name pool $\{Q, M\}$. The asterisk (*) indicates that the model is statistically neutral in the respective interaction. In interaction 1, agents are initially unbiased ($p = 0.116$, see also Fig.~\ref{fig:conv counts}B), based on 10,000 name selections by agents with empty memory. In interaction 2, the convention production probability remains unbiased ($p = 0.110$) when aggregated across equally likely memory configurations. Agents generally adhere to a winning convention but switch to their co-player’s convention following failure. By interaction 3, the dominant memory configurations display a considerable bias towards the strong convention, $M$ ($p<2.2\times10^{-16}$). In stochastic simulations, some agents will inevitably interact with others who have experienced more interactions. These interactions create a bias toward the strong convention, as experienced players are more likely to favor it. Thus, this table provides a conservative estimate of the collective bias emerging for the strong convention.}}
    \label{table: microdynamics}
\end{table}

\begin{table}
\centering
\begin{tabular}{lc}
\hline
\textbf{Model Name}    & \textbf{Model Version} \\
\hline
\textbf{Llama-3-70B-Instruct }               & Meta-Llama-3-70B-Instruct                     \\
\textbf{Llama-3.1-70B-Instruct      }          & Meta-Llama-3.1-70B-Instruct                     \\
\textbf{Claude-3.5-Sonnet}                 & claude-3-5-sonnet-20240620                   \\
\textbf{Llama-2-70b-Chat}                & Meta-Llama-2-70b-Chat                     \\
\hline
\end{tabular}
\caption{Model Names and Versions}
\label{tab:model_versions}
\end{table}



\clearpage 

%

%
%
%
%
%
%


\subsection*{Acknowledgments}
\paragraph*{Funding:}
L.~M.~A. acknowledges the support from the Carlsberg Foundation through the COCOONS project (CF21-0432). 
\paragraph*{Computation:}
We acknowledge City St George's, University of London's Hyperion cluster for computation time.
\paragraph*{Data and materials availability:}
The code and data used in this manuscript are available in the following GitHub repository:
\url{https://github.com/Ariel-Flint-Ashery/AI-norms}, \url{https://doi.org/10.5281/zenodo.14937173}.
\paragraph*{Author contributions:}
A.~F.~A., L.~M.~A., and A.~B. designed the study. A.~F.~A. and L.~M.~A performed the experiments. A.~F.~A. wrote the code for the experiments. A.~F.~A, L.~M.~A., and A.~B. analyzed the data, discussed the results, and contributed to the final manuscript.
\paragraph*{Competing interests:}
There are no competing interests to declare.


\subsection*{Supplementary materials}
Supplementary Text\\
Figs. S1 to S12\\
Tables S1 to S7\\
References \textit{(8, 55-57, 83-\arabic{enumiv})}\\ 


\newpage


\renewcommand{\thefigure}{S\arabic{figure}}
\renewcommand{\thetable}{S\arabic{table}}
\renewcommand{\theequation}{S\arabic{equation}}
\renewcommand{\thepage}{S\arabic{page}}
\setcounter{figure}{0}
\setcounter{table}{0}
\setcounter{equation}{0}
\setcounter{page}{1} 


\begin{center}
\section*{Supplementary Materials for\\ \scititle}

Ariel~Flint~Ashery$^{1}$,
Luca~Maria~Aiello$^{2,3}$,
Andrea~Baronchelli$^{1,4\ast}$,\\

\small$^\ast$Corresponding author. Email: a.baronchelli.work@gmail.com\\

\end{center}

\subsubsection*{This PDF file includes:}
Supplementary Text\\
Figures S1 to S12\\
Tables S1 to S7\\

\newpage


\section*{Supplementary Text}
\subsection*{Microscopic Bias}
\subsubsection*{Statistical Tests for Table~\ref{table: microdynamics}}
To measure the strategy bias in Table~\ref{table: microdynamics}, we assess both the interaction-level bias and the bias within each unique memory configuration. The interaction-level bias is defined as the overall production probability of the strong convention across all possible configurations (per interaction), which we test using an exact Binomial test with a null probability of $p=0.5$, rejecting the null hypothesis if the p-value falls below $0.05$. The results of the tests are reported in the caption of Table 1. 
At the configuration level, we first perform an exact Binomial test, as above, to check whether the model is biased. In all cases, the p-value $p^{\dagger} < 0.05$, confirming that the model's decision is biased towards an extreme. Then, we use bootstrapping by resampling $70\%$ of the observations for each configuration $10,000$ times and measure the proportion of samples showing a stronger bias than the observed value in Table 1. In all cases, we obtain a bootstrapped $p^{\ddagger} > 0.05$, indicating that we cannot reject the hypothesis that the model's underlying bias is more extreme than the observed bias. Results are reported in Table~\ref{tab: bootstrap}.

\subsubsection*{Effect of Fine-Tuning}
We reproduced the set of experiments reported in Table~\ref{table: microdynamics} for \textbf{Llama-3.1-70B-Instruct} (w/ Fine-Tuning) using \textbf{Llama-3.1-70B} (w/o Fine-Tuning), to investigate the effects of instruction-fine-tuning on strategy bias. We ran a 4-bit quantization of the LLM on a single A100 GPU and extracted the next-token probability distribution over the valid convention tokens. The initial, no-memory probability distribution for the same convention choices as in Table~\ref{table: microdynamics} is biased for the pre-trained model, with $p(Q) = 0.525$. To faithfully reproduce any potential biased symmetry breaking, we conduct the experiments using a pair of six-character strings, each consisting of randomly chosen letters and numbers. Most importantly, we repeatedly sampled random pairs until we found a combination that had Jensen-Shannon distance of less than 0.005 from the neutral distribution in the no-memory stage, i.e., an initially unbiased combination. The results are reported in Table~\ref{tab: pretrained-bias}, using the pseudo-labels `A' and `B' for the random conventions. We find that although there is no initial preference for either convention, a strong collective bias already appears by the second interaction. Thus, we cannot say that fine-tuning is responsible for inducing collective bias effects.

\subsection*{Theoretical Model}
\label{SI: minimal NG}

The \emph{naming game} model simulates a population of $N$ agents engaging in pairwise negotiation interactions, demonstrating the emergence of global consensus on conventions through local coordination mechanisms. In the canonical formulation \cite{baronchelli2006sharp}, agents must reach consensus on the name for an object using only local interactions, similar to our experimental framework.
Agents possess internal lexicons with unlimited word capacity (although this is not a necessary initial condition of the model), initially empty. The interaction protocol involves random selection of agent pairs, where the designated speaker transmits a randomly chosen word from their lexicon (or invents a new one if it is empty) to the hearer. If the hearer recognizes the word in their own lexicon, both agents retain only the communicated word, while in case of failure, the hearer incorporates the novel word into their lexicon. The non-equilibrium dynamics of this system exhibit three distinct temporal phases: \emph{(i)} an innovation phase characterized by word creation, \emph{(ii)} a propagation phase involving lexicon reorganization, and \emph{(iii)} a convergence phase culminating in global consensus. In our experimental framework, we set an initial condition whereby agents can only invent new words from a finite word pool of size $W<N$. This condition means that the initial innovation phase is extremely short, as seen in the inset of Fig. 1. This model provides insights into the dynamics of language evolution and convention formation in both human and artificial communication systems.

Fig.~\ref{fig: theoretical production} shows the production probability trajectories of a simulation of the theoretical model with a lexicon of two words ($W=2$).

\subsection*{Characterizing Relative Strength}
We define the relative strength of a convention as its ability to attract the system toward an equilibrium consensus on it, and retain the system in this consensus state. As such, the relative strength of a `strong' convention over its `weak' counterpart is best demonstrated through the latter's ability to overturn consensus on the former. Importantly, we note that the trajectory of the system is sensitive to initial conditions, such that not all memory configurations are observed during the evolution of the system. Fig.~\ref{fig: endgame strategy distribution} shows the complete probability distribution for all memory configurations given that the memory size is $H=5$, while Fig.~\ref{fig: relative strength} provides a characterization of the relative strength of convention pairs in different models in two settings: we evolve the system from a consensus on the strong convention \emph{i)} without perturbation and \emph{ii)} with a perturbation of a single agent committed to the weak convention. We show that the relative strength of a convention with respect to its counterpart can be characterized by the basin of attraction of the steady state on the strong convention. Although \textbf{Llama-3.1-70B-Instruct} agent populations tend to be more exploratory, even in the absence of perturbation, the memory state bounds of their exploration have strong resistance and return the system towards a state of equilibrium around the strong convention steady state. In contrast, while \textbf{Llama-3-70B-Instruct} agent populations do not explore beyond the consensus state once they reach it, they are far more susceptible to perturbations, with some \textit{individual} agents briefly exhibiting memory configurations in which all past interactions within their memory range reflect successful coordination on the weak convention.
Finally, Fig.~\ref{fig: emergence strategy distribution} shows that relative strength also plays a role in convergence, restricting the exploration of of the system in the path towards consensus for \textbf{Llama-3.1-70B-Instruct} population. We therefore propose that the relative strength of a convention can be characterized by properties of its steady-state basin of attraction. The `width' of the basin reflects the range of initial conditions that lead to convergence on the convention. The `pull' represents the stability of the steady state, capturing both its resistance to perturbations and the tendency of trajectories to return after deviations, as well as the rate at which they are drawn toward the steady state.

\subsection*{Prompting}
\label{SI: prompting}
\subsubsection*{Prompt Structure}
The system prompt comprises of three components: \emph{i)} a fixed prompt that outlines the game’s rules, including the payoff structure and the player’s objective, \emph{ii)} a dynamic memory prompt that provides contextual information about the state of play within the player's memory range, and \emph{iii)} an instructional prompt that provides information for how the agent should format its response. We find that agents generally struggle to behave in a manner befitting a partnership game, and often opt for strategies aimed at undermining their co-player's payoff, effectively treating them as an opponent. In practice, this meant that on some occasions the agent would willingly take an action with a negative payoff, in order to harm their co-player's accumulated point tally. 

\subsubsection*{Output Structure}
To extract any meaningful decision from an LLM's output, which may be verbose and unstructured, it is necessary to distinguish between the \textit{reason}, where the agent `shows its working', and the final \textit{decision}. One popular approach to prompting asks the agent to give a final decision at the end of its answer, allowing it to generate a reasoning for the decision before reporting the actual response. This method relies on the assumption that if the LLM is good at composing a well thought-out reasoning, having it spelled out explicitly would guide the choice of the LLM towards better performance due to the LLM's auto-regressive text generation method. Although this approach has shown promising results in a variety of tasks and multi-player games, there is an ongoing debate as to whether LLMs truly possess the ability to reason about the possible actions, beliefs, and intentions of their fellow players~\cite{akata2023playing, ullman2023large, aher2023using}. 

One can argue that even if agents were able to reason, the wording generated to explain the reasoning can potentially express the agent's biases and influence the final decision. As a result, the reason-first, answer-later structure would make it difficult to identify the possible sources of bias within the prompt, especially since the agent's reasoning will only serve to amplify its own bias reflected in the decision.
To strengthen generalization and ensure that the final action decision is influenced solely by the input prompt, we instead ask the agent explicitly to provide an output in the form \textit{answer-first-reason-later}.

\subsubsection*{Example Prompt}
We provide an example of the system and user prompts given to LLM agents in our experiments. We exclude the beginning of text and end of text tokens, which are unique to each model.

\TitledBox{System Prompt}{Context: Player 1 is playing a multi-round partnership game with Player 2 for 100 rounds. At each round, Player 1 and Player 2 simultaneously pick an action from the following values: [\textit{F, J}].
    The payoff that both players get is determined by the following rule:
    
    1. If Players play the SAME action as each other, they will both be REWARDED with payoff 100 points.
    
    2. If Players play DIFFERENT actions to each other, they will both be PUNISHED with payoff -50 points.
    
    The objective of each Player is to maximize their own accumulated point tally, conditional on the behavior of the other player.
    This is the history of choices in past rounds:
    
    \{'round':1, 'Player 1': F, 'Player 2': J, 'payoff': -50\}
    
    .
    
    .
    
    .
    
    It is now round \textit{i}. The current score of Player 1 is \textit{SCORE}. Answer saying which value Player 1 should pick. Please think step by step before making a decision. Remember, examining history explicitly is important. Write your answer using the following format: \{'value': \textit{$<$VALUE\_OF\_PLAYER\_1$>$}; 'reason': \textit{$<$YOUR\_REASON$>$}\}.}

\TitledBox{User Prompt}{Answer saying which action Player 1 should play.}

\subsubsection*{Meta-Prompting}
When LLMs are used to solve tasks where some form of ground truth is defined, such as classification or regression, the effect of prompt variations on the quality of a model's outputs can be measured on downstream performance~\cite{lester2021power}. However, that is not possible in generative tasks where a notion of error is undefined. Specifically in the naming game, any generated output is plausible, as long as it is within the set of allowed symbols. This ambiguity makes it difficult to assess whether the LLM's outputs reflect a proper semantic understanding of the task's rules or are merely products of statistical `hallucinations'~\cite{xu2024hallucination}. To partially address this issue, we rely on a meta-prompting technique to measure the LLMs' level of comprehension of the given prompt~\cite{fontana2024nicer}. This technique provides the LLM with the prompt, and then asks three types of \emph{prompt comprehension questions} about: interaction rules, chronological sequence of actions in the history, and payoff statistics (Table~\ref{tab:meta-prompting}).

\begin{figure}[!h]
\centering
\includegraphics[width=0.8\textwidth]{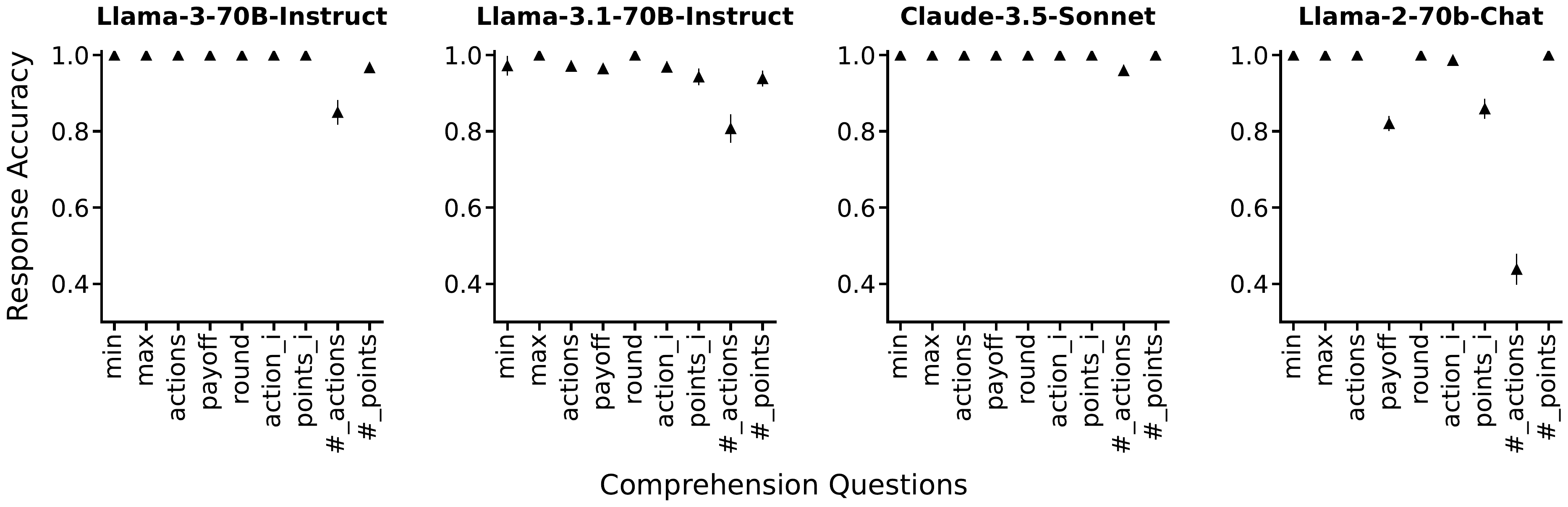}
\caption{\textbf{Meta-prompting results} Accuracy of the model responses to the prompt comprehension questions defined in Table~\ref{tab:meta-prompting}. We selected 8 agents from a single run (5 agents for \textbf{Llama-3-70B-Instruct}), and recovered their game record. We replayed the game using the memory length used in the simulated run ($H=5$), posing the comprehension questions at each interaction. These runs provide approximately $100$ test interactions for each model.}
\label{fig: meta prompting}
\end{figure}

\begin{figure}[h]
\centering
        \includegraphics[width=0.6\linewidth]{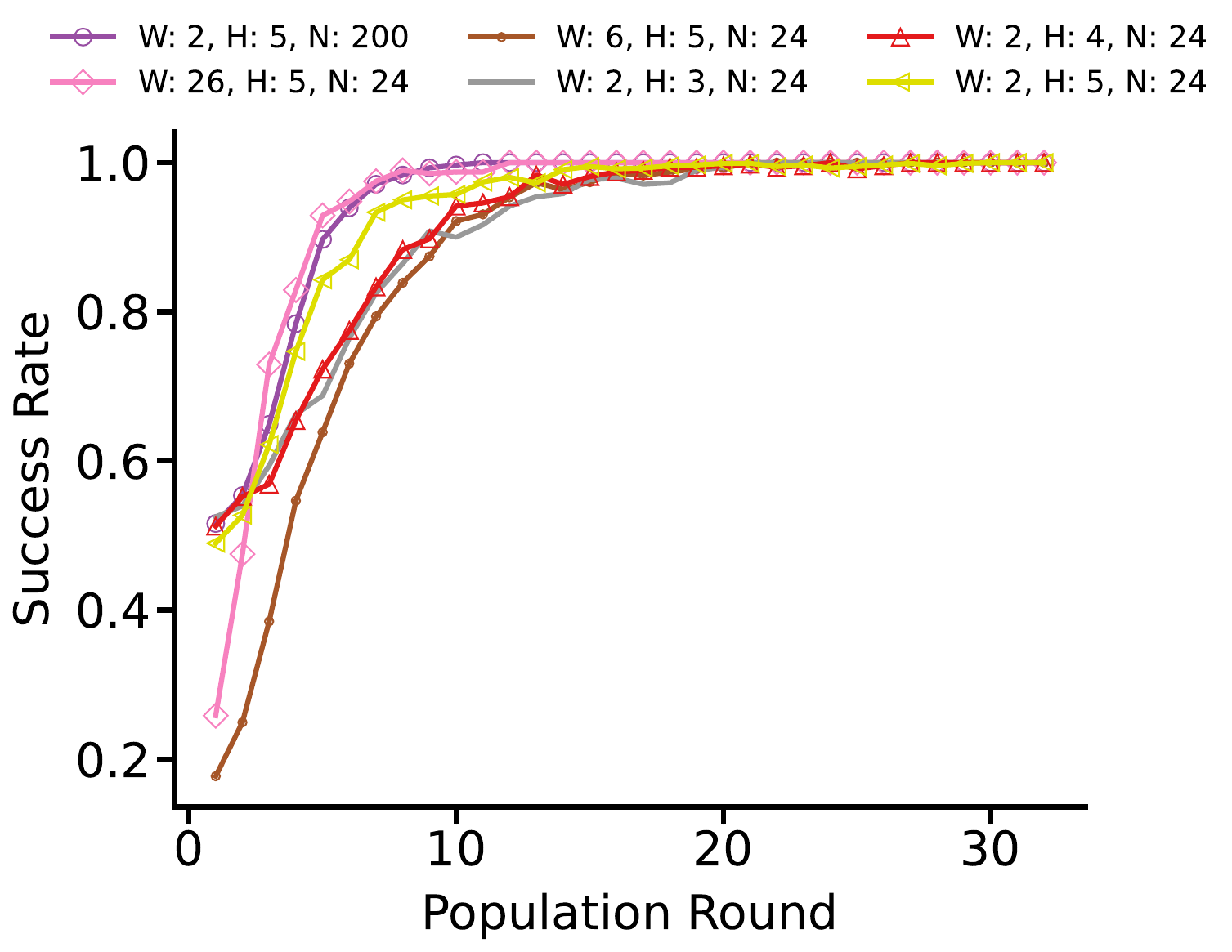}
        \caption{\textbf{Robustness of the Spontaneous emergence of conventions.} We show that the spontaneous emergence of conventions holds for a variety of simulation parameters, using populations of \textbf{Llama-3-70B-Instruct} agents. $W=26$ indicates a name pool which uses the entire Latin alphabet, $W=6$ is the name pool $\{Q, M, F, J, X, Y\}$, and $W=2$ is the name pool $\{Q, M\}.$}
\label{fig: conv robustness}
\end{figure}

\begin{figure}[h]
\centering
        \includegraphics[width=0.6\linewidth]{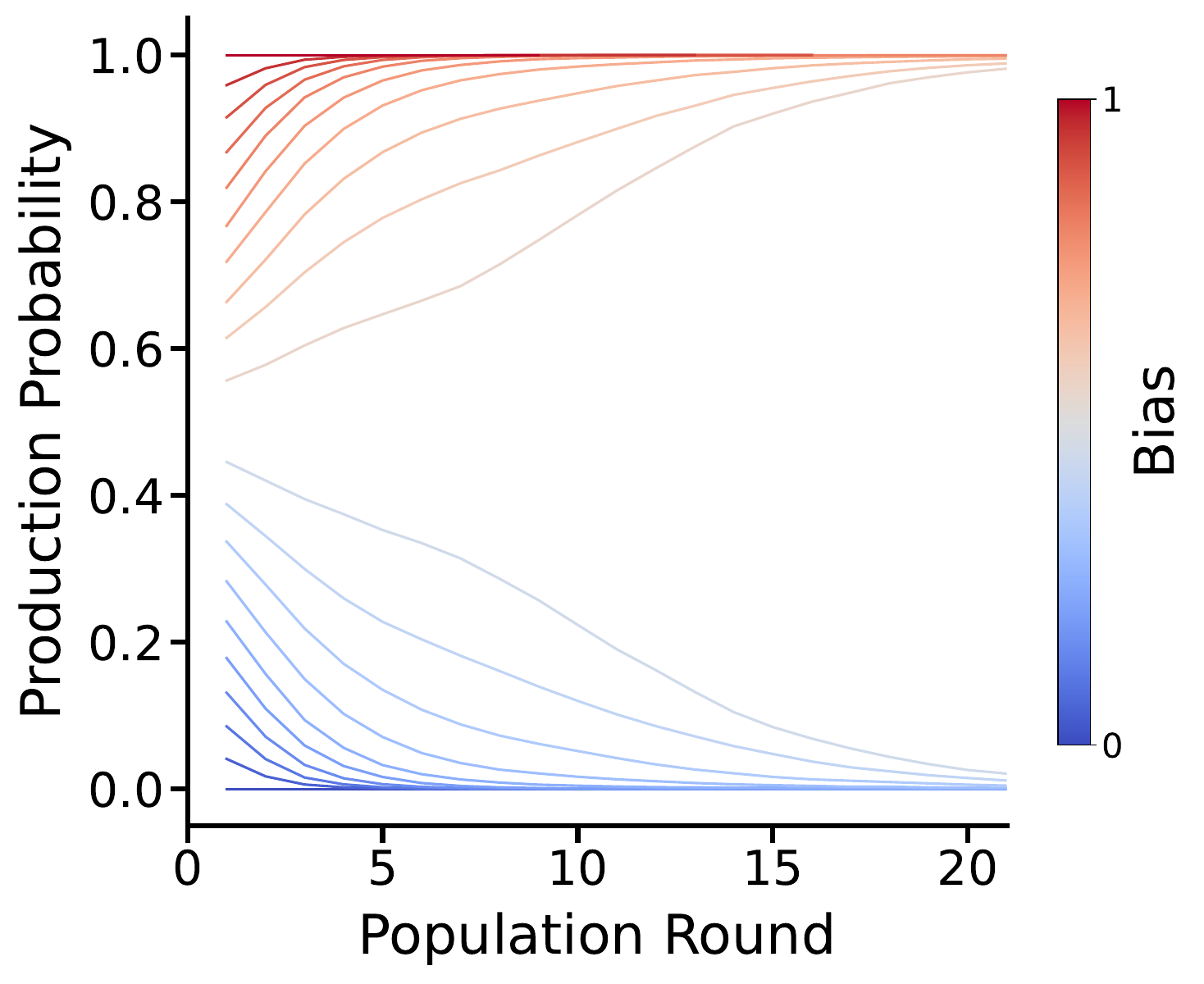}
        \caption{\textbf{Production probability trajectories in the theoretical minimal naming game with $W=2$.} Agents can only choose names from the pool \{0,1\}. We simulated 10,000 runs of the minimal naming game in a population of 24 agents. We show biased trajectories (in probability increments of 0.05) towards choosing the name `1'. The bias corresponds to the probability of choosing the name '1' when an agent has the option of producing either name, such that a bias of 1 (0) corresponds to agents that will only choose name `1' (`0'). We note that as the bias increases, the convergence speed increases. Crucially, even a small bias towards a certain name leads to inevitable global convergence on that name.}
    \label{fig: theoretical production}
\end{figure}

\begin{figure}[h]
\centering
        \includegraphics[width=.6\linewidth]{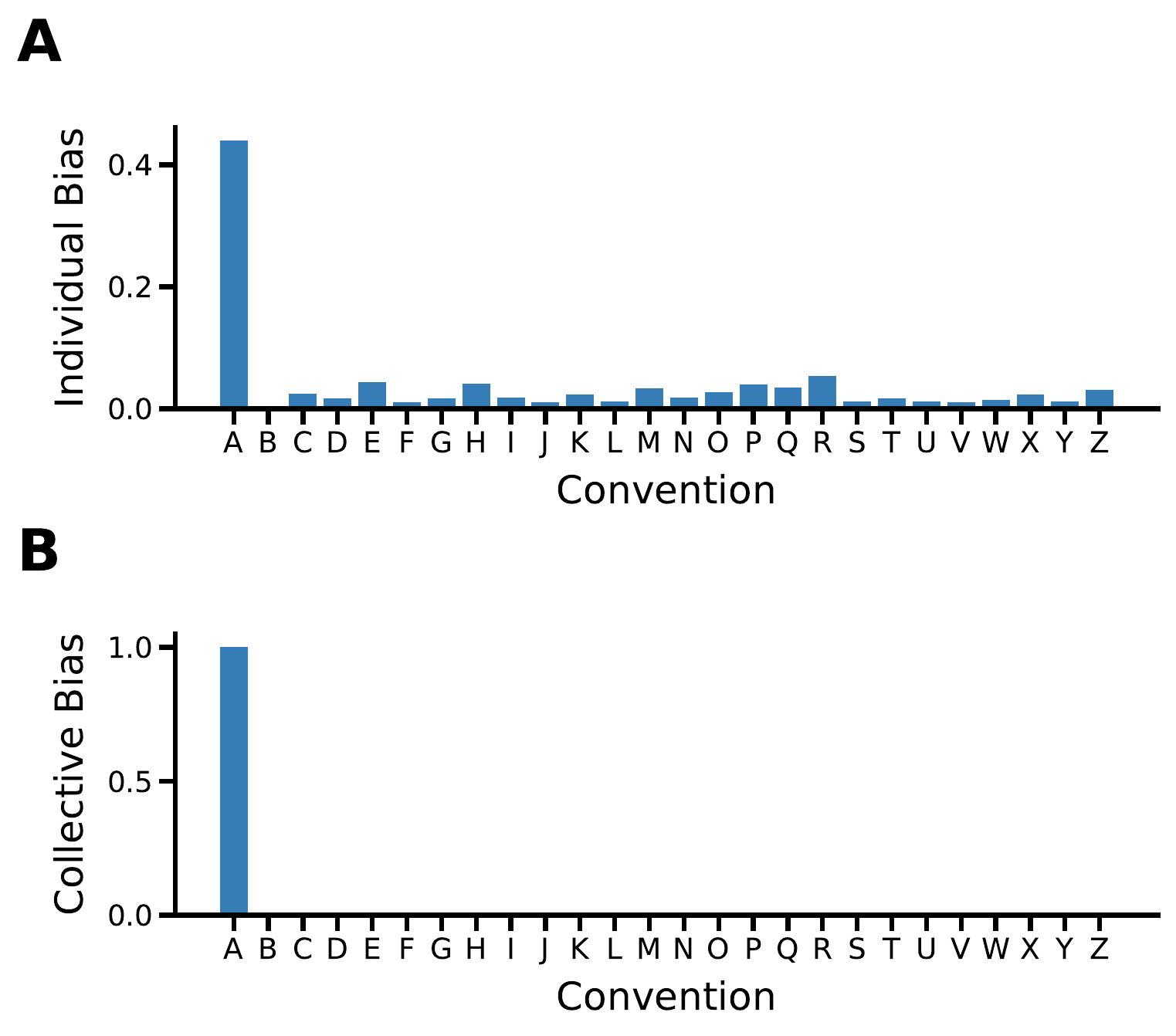}
        \caption{\textbf{(A) Individual and (B) Collective bias in convention selection with $W=26$, the entire Latin Alphabet.} Agents favor the convention `A' over all others \textit{a priori}, resulting in collective consensus on this convention. Individual bias shows the convention production probability from 480 samples using \textbf{Llama-3-70B-Instruct} agents, where agents have empty memory. Collective bias shows the proportion of consensus conventions from 20 simulations.}
    \label{fig: latin emergence}
\end{figure}

\begin{figure}[h]
\centering
        \includegraphics[width=0.6\linewidth]{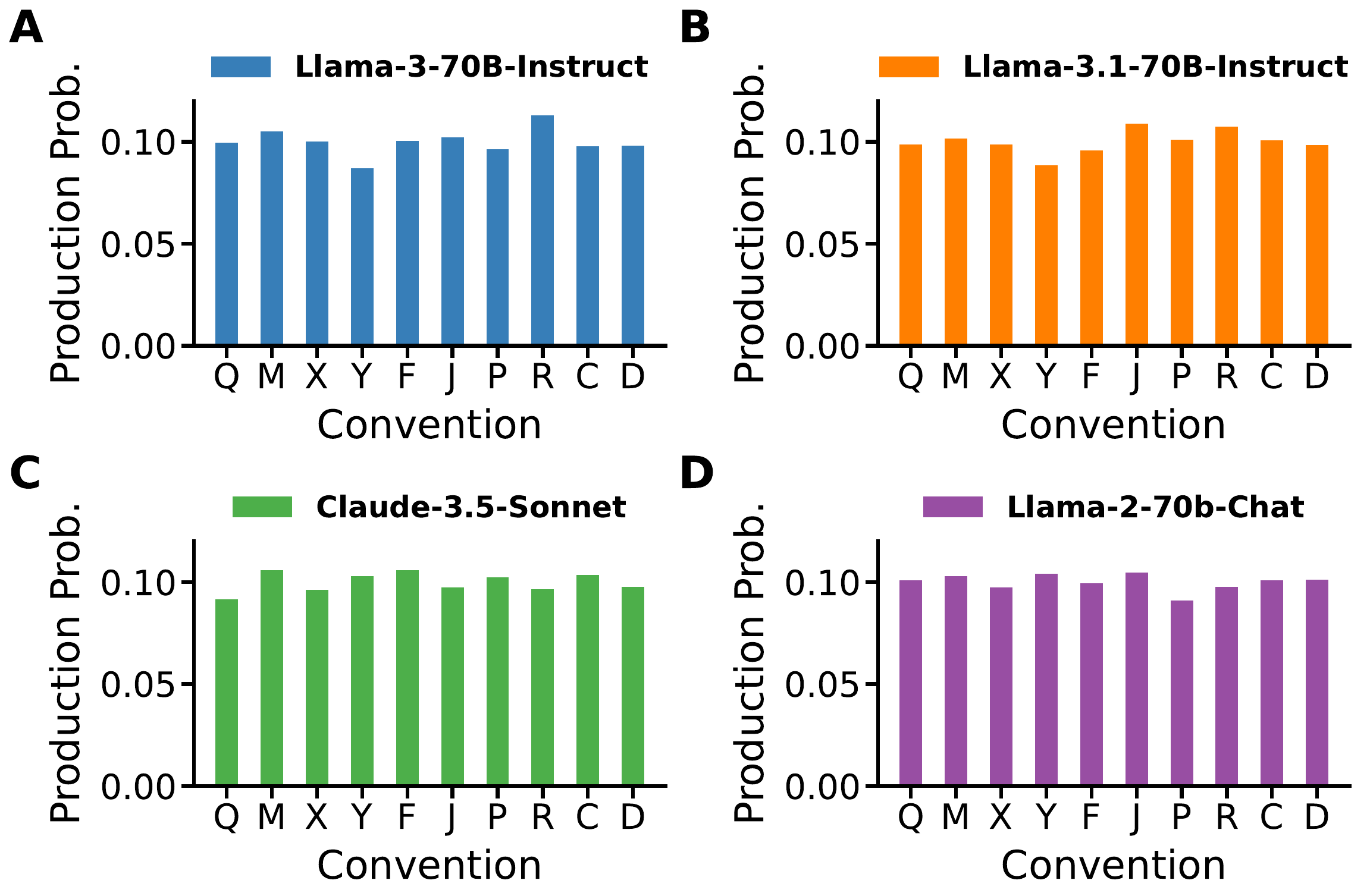}
        \caption{\textbf{Individual bias in conventions selection with $W=10$.} The production probability of each convention when agents have no prior memory, for the LLM agents indicated in the legend. For (\textbf{A}-\textbf{D}), we generated 15,000, 10,000, 4,500, and 10,000 samples. We performed a chi-squared hypothesis test to see whether the agents are biased, and calculated the following p-values: $p$ = $<$ 0.001, 0.001, 0.100, 0.410. These p-values indicate that \textbf{Llama-2-70b-Chat} and \textbf{Claude-3.5-Sonnet} are unbiased across conventions in this name pool (at the 5\% significance level), whereas the \textbf{Llama-3/3.1-70B-Instruct} models exhibit significant skew.}
    \label{fig: ten individual bias}
\end{figure}

\begin{figure}[h]
\centering
\includegraphics[width=0.6\linewidth]{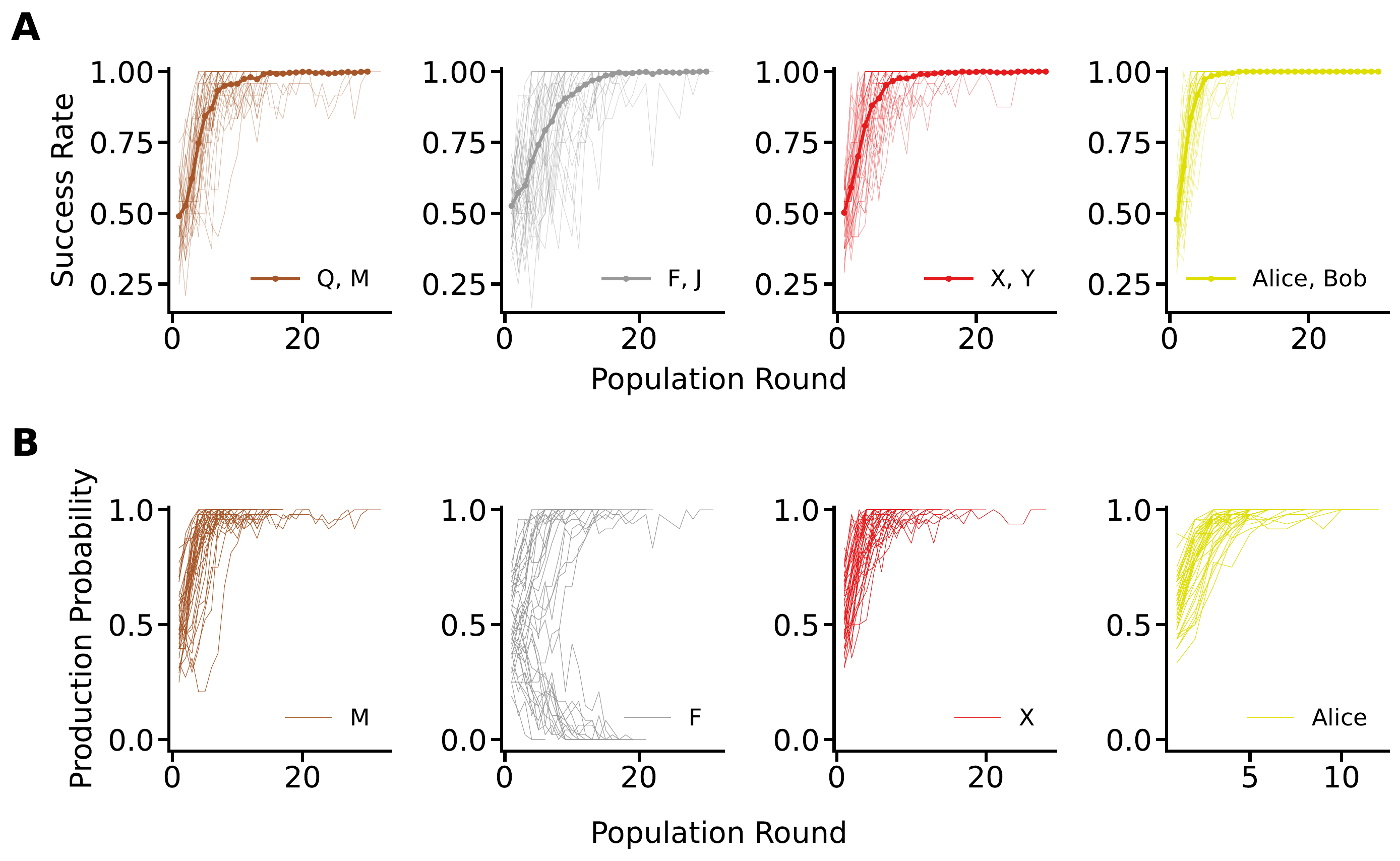}
\caption{\textbf{The Spontaneous emergence of conventions for $W=2$.} We present individual (faint lines) and average (thick lines) trajectories of a population of $N=24$ \textbf{Llama-3-70B-Instruct} agents with memory length $H=5$ for four different name pools. For each name pool, we show \textbf{(A)} the success rate, and \textbf{(B)} the production probability of the strong convention (as indicated by the legend). All name pools resulted in a collective bias on a particular convention ($\{Q, M\}: M$, 40/40 runs;  $\{F, J\}: F$, 24/40 runs; $\{X, Y\}: X$, 40/40 runs; $\{Alice, Bob\}: Alice$, 40/40 runs).}
\label{fig: binary emergence}
\end{figure}

\begin{figure}[!h]
\centering
\includegraphics[width=0.6\linewidth]{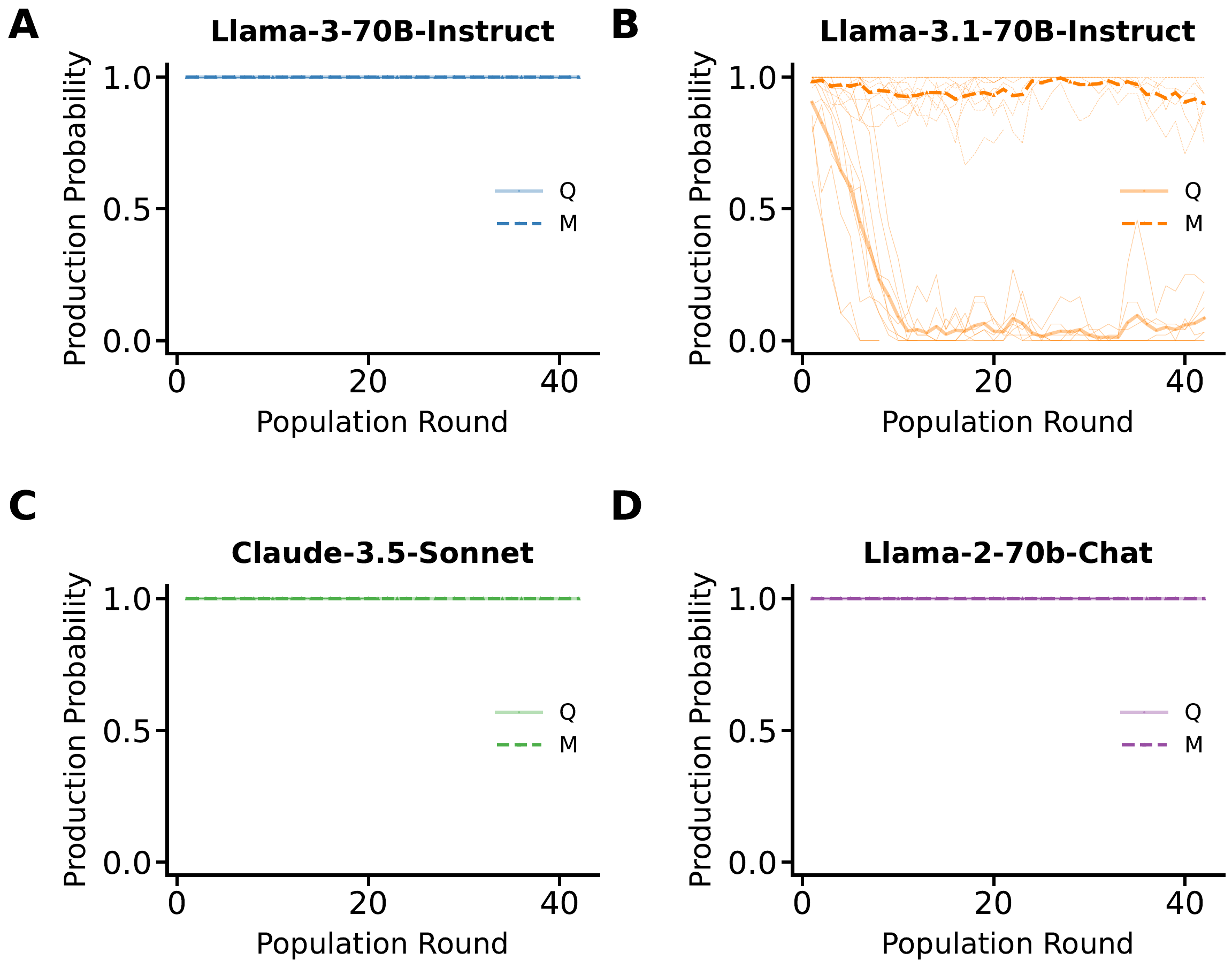}
\caption{\textbf{Stability of consensus conventions.} We test the stability of each model in the setting $W=2$, with the possible conventions shown in the legend. For each convention, we begin with a population of $N=24$ agents ($N=48$ for \textbf{Llama-3-70B-Instruct}), where every agent has only the respective convention in memory. We then allow the population to naturally evolve, and measure the production probability of this convention. We simulate the following number of runs (\textbf{A-D}): 4, 10, 3, 5. Faint lines show the trajectories of individual runs. We show that for all models apart from \textbf{Llama-3.1-70B-Instruct}, the population remains entirely stable at its initial consensus state. If \textbf{Llama-3.1-70B-Instruct} is initialized with consensus on the weak convention ($Q$), the population immediately switches to the alternative convention ($M$, the strong convention). Here, the strong convention remains stable, with some minor fluctuations. The instability of the weak convention is also observed in our study of the committed minority required to flip a majority consensus (see Fig.~\ref{fig: tipping points}), where we see that $M$ acts as a strong attractor state that can only be overcome by a large enough committed minority on the weak convention, $Q$.}
\label{fig: stability robustness}
\end{figure}

\begin{figure}[!h]
\centering
\includegraphics[width=0.6\linewidth]{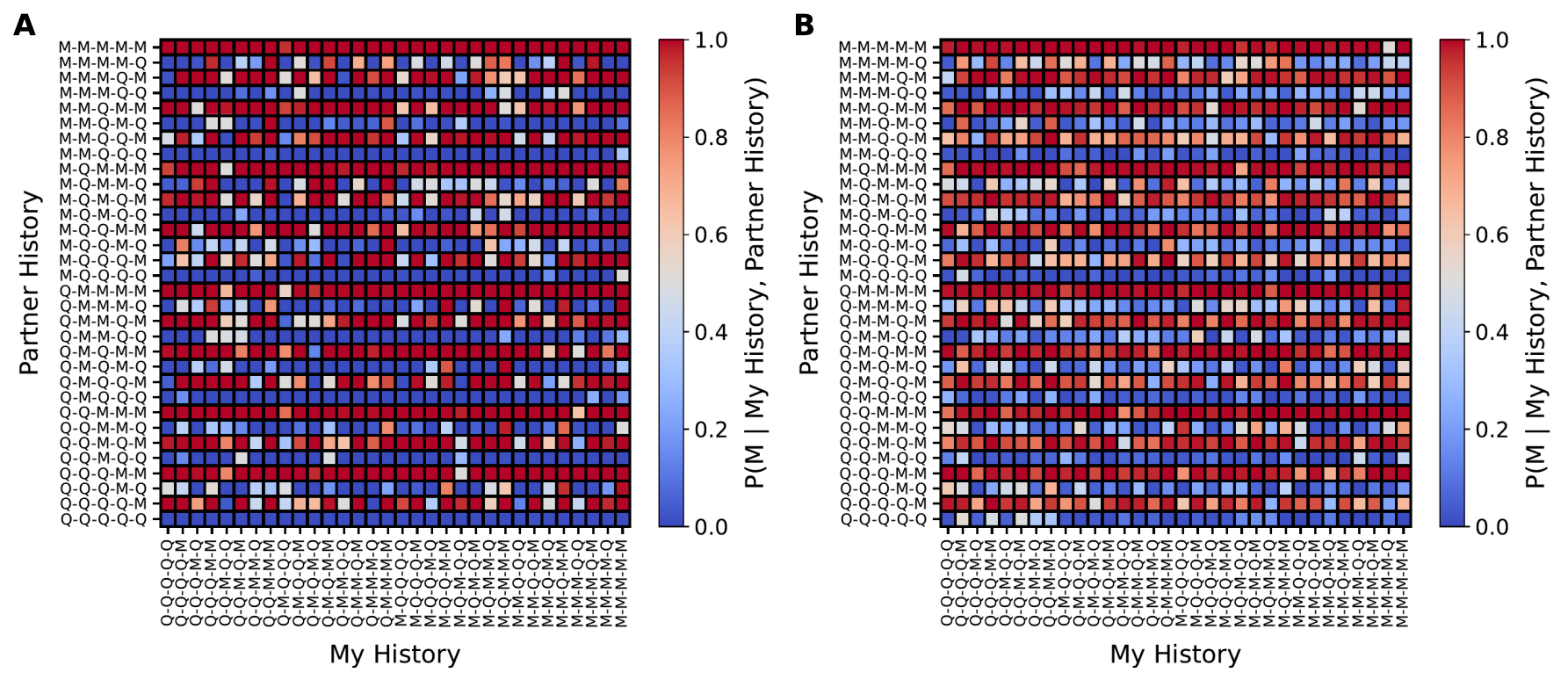}
\caption{\textbf{Distribution of strategies across memory states.} We `open’ the black-box to show the entire probability distribution across memory states for $H=5$ for (\textbf{A}) \textbf{Llama-3-70B-Instruct}, and (\textbf{B}) \textbf{Llama-3.1-70B-Instruct} agents. Each tile represents a joint history that comprises the agent’s memory state. The color of the tile indicates the probability that the agent will choose the convention $M$ in the next interaction, which remains fixed for each memory state at all times.
}
\label{fig: endgame strategy distribution}
\end{figure}

\begin{figure}[!h]
\centering
\includegraphics[width=0.6\linewidth]{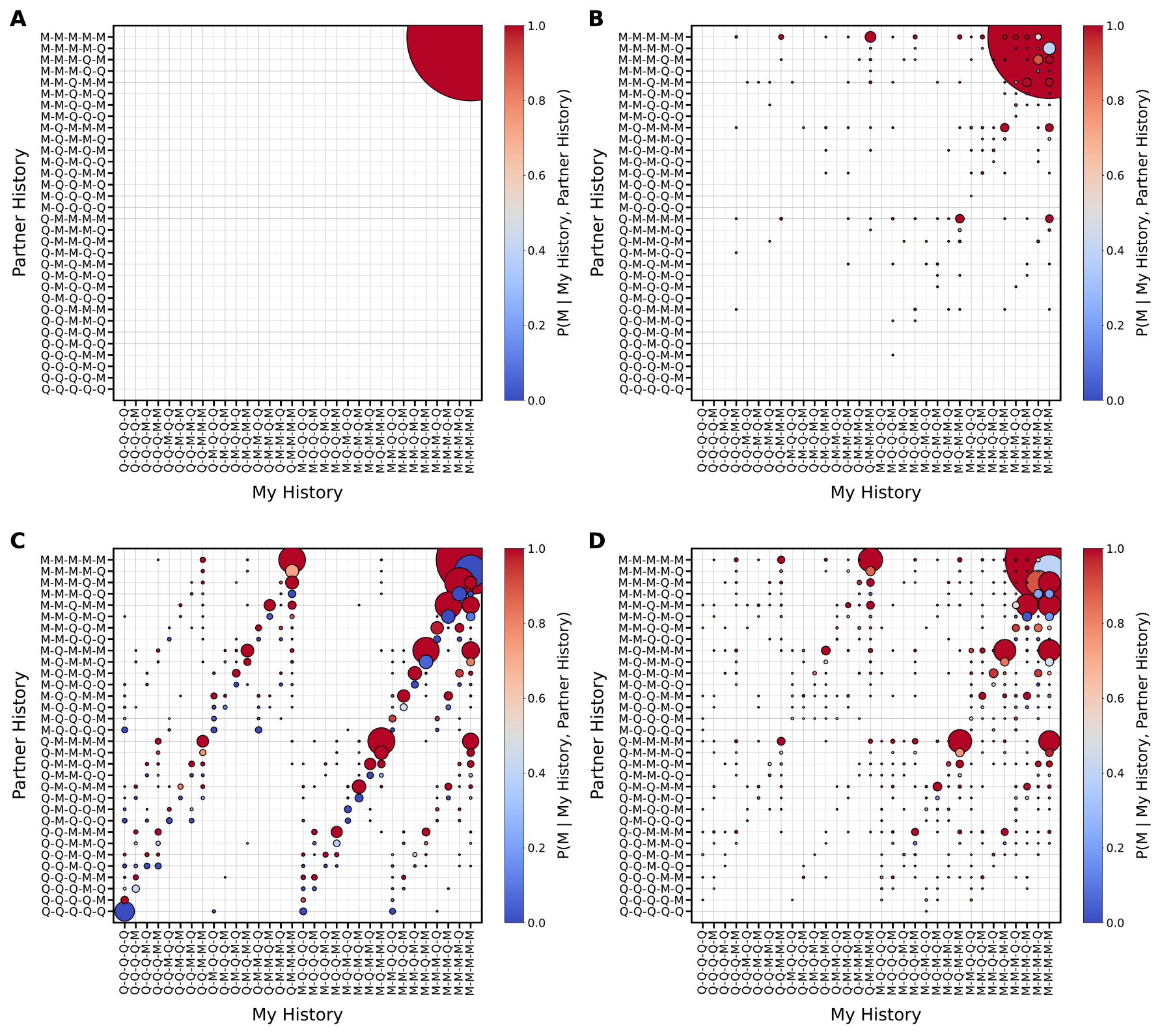}
\caption{\textbf{Microscopic characterization of memory-dependent strategy bias in real-game dynamics.} The matrices depict the actual strategies of agents for memory states of size $H=5$ during the game. Given a memory state defined by a joint history, the color of the circles corresponds to the probability of generating the dominant convention ($M$), which is exact and remains constant between runs, regardless of the system state. The size of the circles corresponds to the average frequency of the memory state across 10 simulated runs. We study two settings: starting from an initial consensus on the dominant convention ($M$) we let the system (\textbf{A-B)} evolve naturally, and (\textbf{C-D}) evolve with perturbation of a single agent committed on the other convention ($Q$), both for 100 population rounds with $N=24$. The panels on the left (right) show the dynamics for a population of \textbf{Llama-3-70B-Instruct} (\textbf{Llama-3.1-70B-Instruct}) agents.}
\label{fig: relative strength}
\end{figure}

\begin{figure}[!h]
\centering
\includegraphics[width=0.6\linewidth]{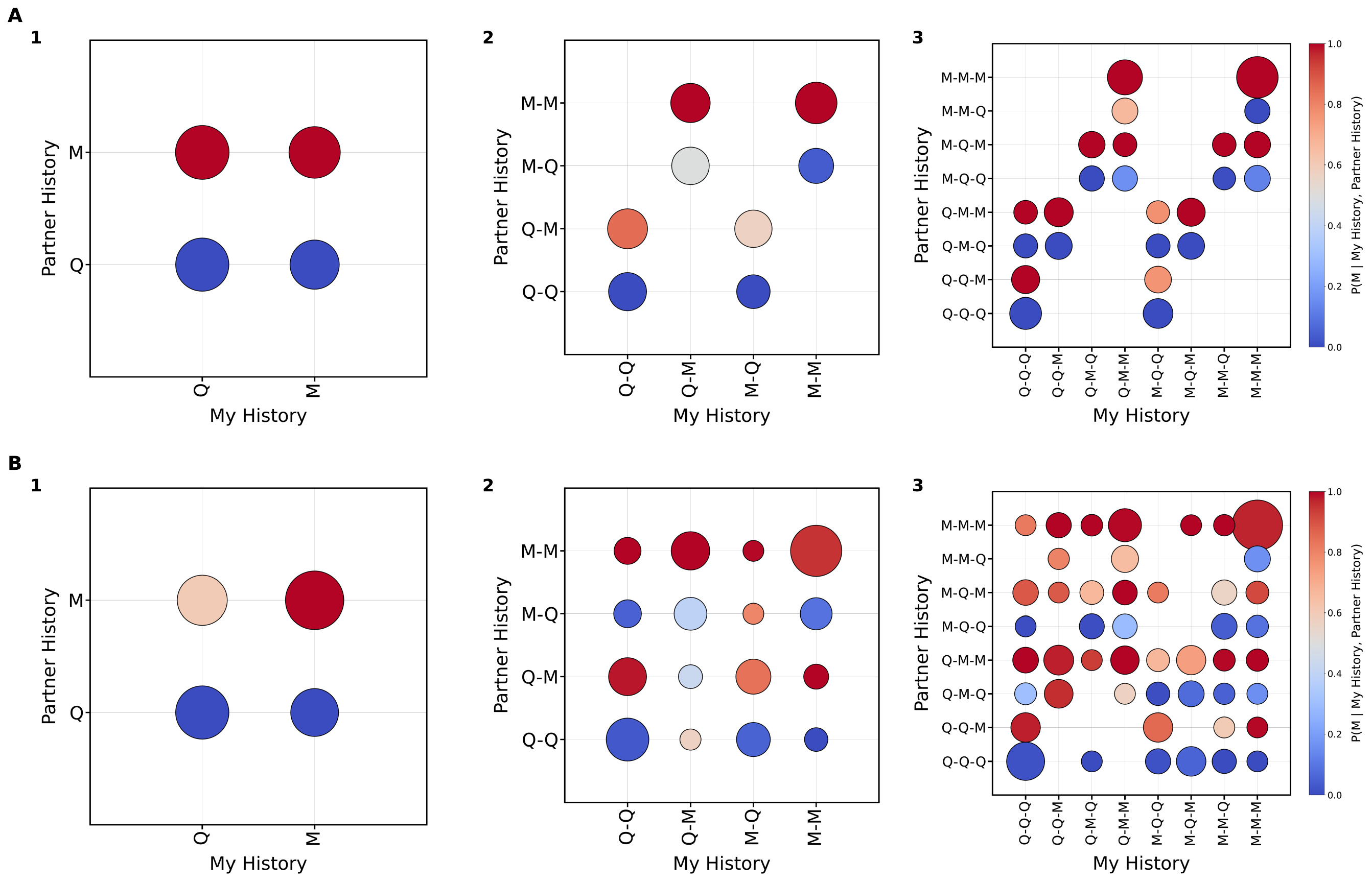}
\caption{\textbf{Distribution of strategies across memory configurations during spontaneous emergence.} We study the emergence of consensus in \textbf{Llama-3-70B-Instruct} (top row, \textbf{A1-3}) and \textbf{Llama-3.1-70B-Instruct} (bottom row, \textbf{B1-3}), with the name pool $\{M, Q\}$. Given a memory state defined by a joint history, the color of the circles corresponds to the probability of generating the dominant convention ($M$), which is exact and remains constant between runs, regardless of the system state. The size of the circles corresponds to the average frequency of the memory state across 40 simulated runs. Note that without observing the frequency of the memory states during game dynamics, predicting a final consensus based on the probability distribution in round 2 (\textbf{B1}) for \textbf{Llama-3.1-70B-Instruct} would result in an incorrect prediction: initially, each memory configuration is equally likely, so $p(Q) > p(M)$ on aggregate. However, we know that the bias flips as the game unfolds, and the population eventually converges on $M$.
}
\label{fig: emergence strategy distribution}
\end{figure}

To assess the LLMs' proficiency in responding to meta-prompting questions, we randomly selected a group of agents from a real stochastic simulation of the naming game. For each agent, we used its history to replay each of its previous interactions, using the same memory length it had in the simulation. Retracing every interaction, including the agent's memory at the time, we ask the agent all possible comprehension questions. Note that certain questions that rely on memory cannot be asked in the first interaction. We pose the questions at each interaction and compute the average accuracy of the LLMs' responses across all interactions for all agents. Overall, all models exhibit a good level of prompt comprehension, with response accuracy nearly always above $0.8$ and most often close to $1$ (Fig.~\ref{fig: meta prompting}). The only model that went below $0.8$ in any metric is \textbf{Llama-2-70b-Chat}, which showed relatively poor accuracy in counting the number of times it played a convention within memory range. In many cases, this agent confused the ID of the player it was asked to consider, or it answered how many times a convention has been observed in total, across both players. Here, it is also worth noting that LLMs from \textbf{Llama-2-70b-Chat}'s generation generally struggled with counting tasks~\cite{huang2023large, fontana2024nicer}.

\subsubsection*{Alternative Prompts}
\begin{figure}[!h]
\centering
\includegraphics[width=0.6\linewidth]{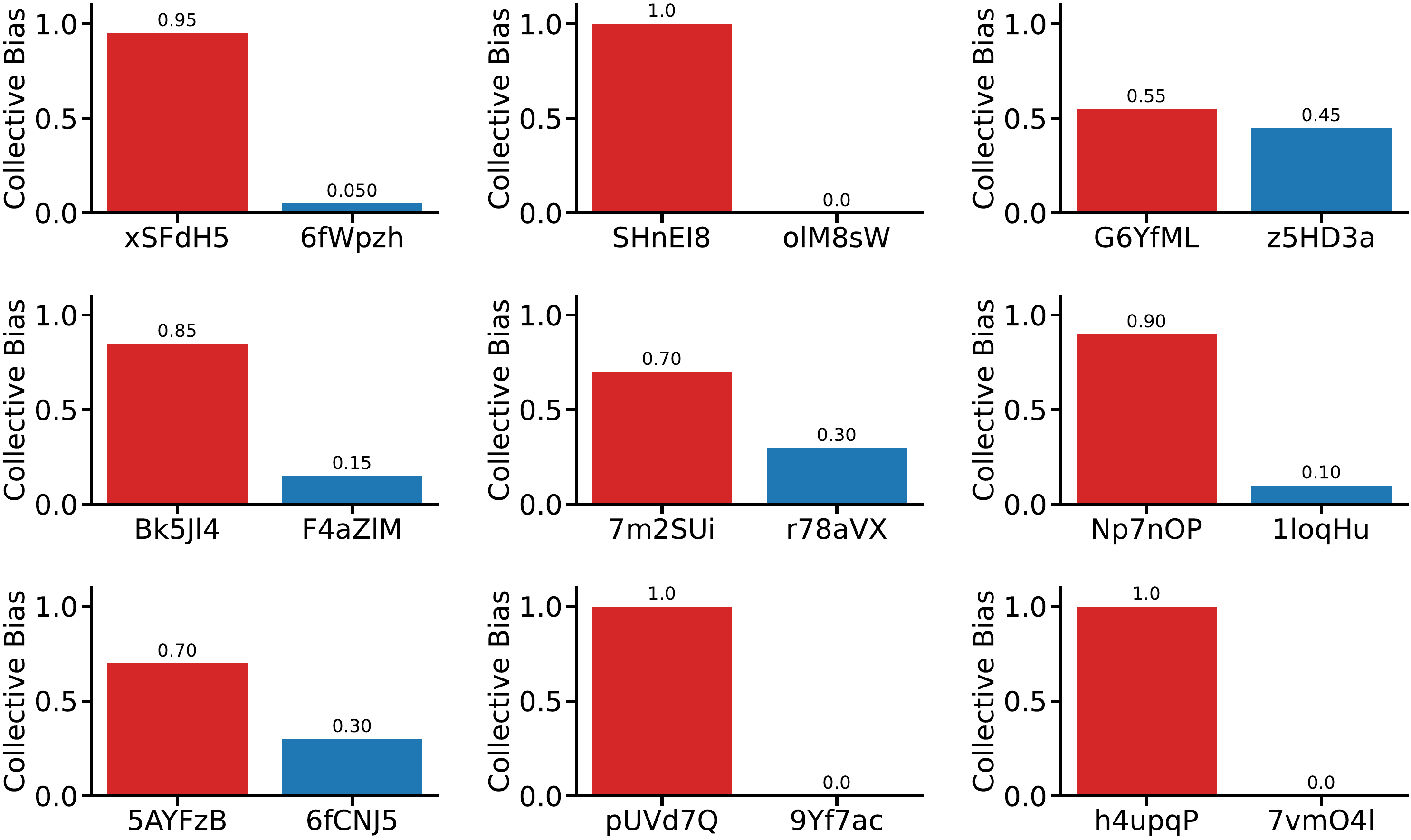}
\caption{\textbf{Collective bias in a population of N=24 \textbf{Llama-3-70B-Instruct} using random convention pairs}. We simulated 20 runs for each pair using the same prompt in the main results. All pairs are initially unbiased, with a Jensen-Shannon distance $<0.005$ from the neutral distribution. 
}
\label{fig: random strings}
\end{figure}

\begin{figure}[!h]
\centering
\includegraphics[width=0.6\linewidth]{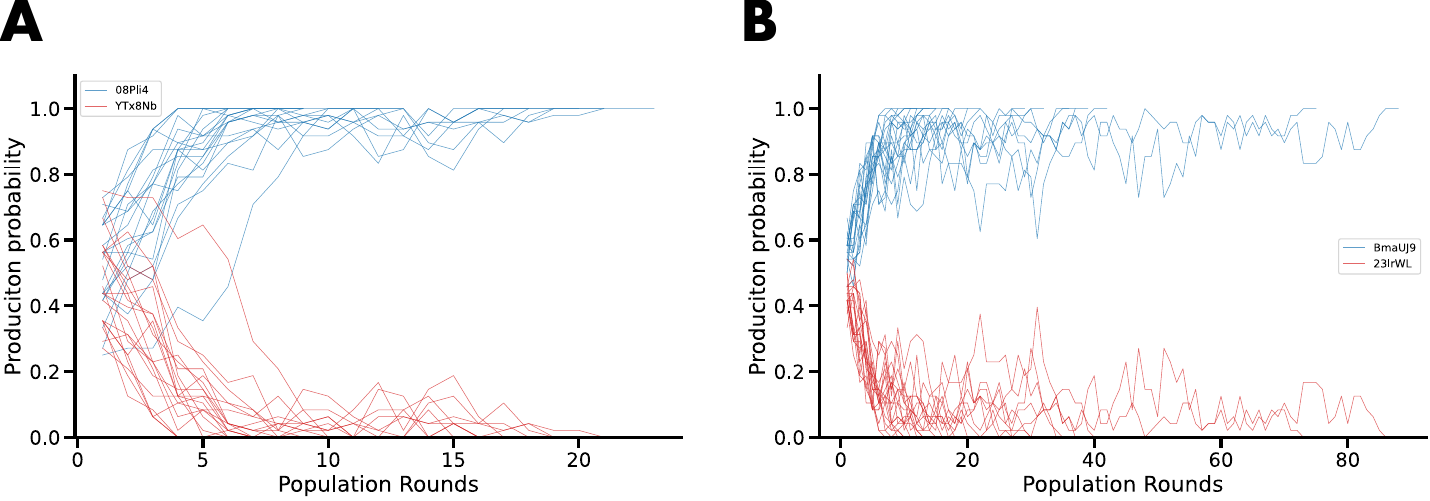}
\caption{\textbf{Collective bias in a population of N=24 \textbf{Llama-3-70B-Instruct} agents using alternative prompts}. We tested the experimental framework using a new prompt template, namely (\textbf{A}) a narrative prompt, and (\textbf{B}) a bullet-point style prompt. For both prompts, we used a name pool of size $W=2$, with random strings such that the initial, no-memory probability distribution over conventions had a Jensen-Shannon distance $<0.005$ from the neutral distribution.  
}
\label{fig: alternative prompts}
\end{figure}

To reduce the risks from bias from previously seen examples in the training data, we test the experimental framework with two additional system prompt templates and using name pools with conventions made up of random strings. The prompt variations are \emph{i)} a factual bullet-point style, and \emph{ii)} an expository, narrative prompt. For the name pools, we generate strings using six characters chosen at random from a set of numerical digits, lowercase and uppercase letters from the Latin alphabet. We only select a name pool if its no-memory probability distribution over conventions has a Jensen-Shannon distance from the neutral distribution less than 0.005, to ensure there is no initial bias. As shown in Figs.~\ref{fig: random strings} and ~\ref{fig: alternative prompts}, biased consensus emerges across all the variations.

In considering whether bias in the prompt arises from the presence of the experimental framework in the training data, we note that the game’s population dynamics and the symmetry in payoffs across action labels ensure that no predetermined optimal strategy exists---even if a player were aware of global information. Convergence occurs purely at the local level, driven by repeated random interactions rather than a global rule that dictates strategic behavior. While individual agents may exhibit micro-level adaptation, adjusting their choices in response to patterns in their co-players’ actions, these adjustments remain localized and do not necessarily lead to a predictable global bias. This contrasts with games like the Ultimatum Game, where systematic behavioral responses shape equilibrium strategies. Similarly, unlike the Prisoner’s Dilemma, where payoff asymmetries guide decision-making, the payoff symmetry in our game framework prevents any clear incentive toward one action over another. As a result, the game’s long-term trajectory remains emergent and stochastic, and only a higher-order metanorm---beyond the direct game mechanics---could impose a systematic bias or predictable pattern.

\TitledBox{Narrative System Prompt}{Context: Player 1 is playing a multi-round partnership game with Player 2 for 100 rounds. At each round, Player 1 and Player 2 make simultaneous action choices between the actions \textit{$<$ACTION\_LABEL\_1$>$} and \textit{$<$ACTION\_LABEL\_1$>$}. The outcome of each round depends on the combination of choices made by the Players in that round. If both Players choose the same action, then the round is a success and both players receive a payoff of +100 points. However, if the Players choose different actions to each other, then the round is a failure and both Players receive a payoff of -50 points. The objective of each Player is to maximise their own accumulated points total. In round 1, Player 1 chose action \textit{$<$ACTION\_OF\_PLAYER\_1$>$} and Player 2 chose action \textit{$<$ACTION\_OF\_PLAYER\_2$>$}; Player 1 received a payoff of \textit{$<$PAYOFF$>$} points and Player 2 received a payoff of \textit{$<$PAYOFF$>$} points... It is now round $i$. The current score of Player 1 is \textit{$<$SCORE\_OF\_PLAYER\_1$>$}. Answer saying which action Player 1 should pick. Please think step by step before making a decision. Remember, examining history explicitly is important. Write your answer using the following format: \{'action': \textit{$<$ACTION\_OF\_PLAYER\_1$>$}; 'reason': \textit{$<$YOUR\_REASON$>$}\}.}

\TitledBox{Bullet-point System Prompt}{Context: Player 1 and Player 2 are playing a repeated 2-player game.

 **Game Setup:** 
 
• Players: Player 1 and Player 2. 

• Actions available to both players: \textit{$<$ACTION\_LABEL\_1$>$} and \textit{$<$ACTION\_LABEL\_1$>$}.

• Players make their choices simultaneously.

 **Payoff Matrix:** 
 
• If Player 1 chooses \textit{$<$ACTION\_LABEL\_1$>$} and Player 2 chooses \textit{$<$ACTION\_LABEL\_1$>$}, then: 

- Player 1 receives +100 points.

- Player 2 receives +100 points.

• If Player 1 chooses \textit{$<$ACTION\_LABEL\_1$>$} and Player 2 chooses \textit{$<$ACTION\_LABEL\_2$>$}, then:

- Player 1 loses -50 points. 

- Player 2 loses -50 points. 

• If Player 1 chooses \textit{$<$ACTION\_LABEL\_2$>$} and Player 2 chooses \textit{$<$ACTION\_LABEL\_2$>$}, then: 

- Player 1 receives +100 points. 

- Player 2 receives +100 points. 

• If Player 1 chooses \textit{$<$ACTION\_LABEL\_2$>$} and Player 2 chooses \textit{$<$ACTION\_LABEL\_1$>$}, then: 

- Player 1 loses -50 points. 

- Player 2 loses -50 points.

 **Player objectives:** 

• The objective of each player is to maximize their own accumulated point tally.}

\TitledBox{Bullet-point System Prompt (Continued)}{**Game History:**

*Round 1:*

• Action choices: 

- Player 1: \textit{$<$ACTION\_OF\_PLAYER\_1$>$}.

- Player 2: \textit{$<$ACTION\_OF\_PLAYER\_2$>$}.

• Payoff: 

- Player 1: \textit{$<$PAYOFF$>$}.

- Player 2: \textit{$<$PAYOFF$>$}.

...

**Next Round:**

It is now round \textit{i}. The current score of Player 1 is \textit{$<$SCORE\_OF\_PLAYER\_1$>$}. You must now decide which action Player 1 should take in this round. Use previous rounds as additional guidance. Remember, the outcome of Player 1's action choice also depends on Player 2's action choice. Write your answer using the following format: \{'action': \textit{$<$ACTION\_OF\_PLAYER\_1$>$}; 'reason': \textit{$<$YOUR\_REASON$>$}\}.}



\newpage

\newpage

\newpage

\newpage

\newpage

\newpage

\newpage

\newpage

\newpage

\newpage

\newpage

\newpage

\newpage

\begin{table}[!h]
\centering
\begin{tabular}{lcc}
\hline
\textbf{Model Name}    & \textbf{Strong Convention} & \textbf{Weak Convention} \\
\hline
\textbf{Llama-3-70B-Instruct}         &         2435                  &      2565                     \\
\textbf{Llama-3.1-70B-Instruct}         &             5079              &      4921                     \\
\textbf{Claude-3.5-Sonnet}        &              5016             &         4984                  \\
\textbf{Llama-2-70b-Chat}           &                  5010         &     4090                      \\
\hline
\end{tabular}
\caption{\textbf{Raw individual bias.} \textmd{Data shown for the left panel in Fig.~\ref{fig:conv counts}B. Values indicate the counts of strong and weak productions in the case $W=2$ for individual agents, showing their preferences \textit{a priori}, when agents are initialized with empty memory.}}
\label{tab:2A data}
\end{table}

\newpage

\begin{table}[!h]
\centering
\begin{tabular}{lcc}
\hline
\textbf{Model Name}    & \textbf{Strong Convention} & \textbf{Weak Convention} \\
\hline
\textbf{Llama-3-70B-Instruct}          &             40              &        0                   \\
\textbf{Llama-3.1-70B-Instruct}        &               40            &       0                    \\ 
\textbf{Claude-3.5-Sonnet}         &                 26          &       14                    \\
\textbf{Llama-2-70b-Chat}          &              36             &        4                   \\
\hline
\end{tabular}
\caption{\textbf{Raw collective bias.} \textmd{Data shown for the right panel in Fig.~\ref{fig:conv counts}B. Values indicate the count of consensus states on the strong and weak conventions after the population converged. For each model, we conducted 40 trial runs. All models had memory length $H=5$.}}
\label{tab:2B data}
\end{table}

\newpage

\begin{table}[!h]
\centering
\begin{tabular}{lcc}
\hline
\textbf{Model Name}    & \textbf{Strong Convention} & \textbf{Weak Convention} \\
\hline
\textbf{Llama-3-70B-Instruct}          &           6                &       1                    \\
\textbf{Llama-3.1-70B-Instruct}         &            10               &       0                    \\
\textbf{Claude-3.5-Sonnet}       &             5              &         5                  \\
\textbf{Llama-2-70b-Chat}         &              16             &         11                  \\
\hline
\end{tabular}
\caption{\textbf{Raw Critical mass values.} \textmd{Data shown for Fig.~\ref{fig: tipping points}. The reported values corresponds to the number of agents required to overturn a majority consensus on the convention: $M$ (strong Convention), and $Q$ (Weak Convention). Population size (top to bottom), $N= 48, 24, 24, 24$}. We performed 10 runs for each experiment, apart from \textbf{Llama-3-70B-Instruct}, which shows 3 runs.}
\label{tab:3B data}
\end{table}

\newpage

\begin{table}[!h]
\centering
\begin{tabular}{lcc}
\hline
\textbf{Parameter} & \textbf{Value} \\ 
\hline
Temperature & 0.5 \\ 
Top-K       & 10  \\ 
Max Tokens  & 6   \\ 
\hline
\end{tabular}
\caption{Model Parameters.}
\label{table:llm_parameters}
\end{table}

\newpage

\begin{table}[!h]
\centering
\includegraphics[width=.6\linewidth]{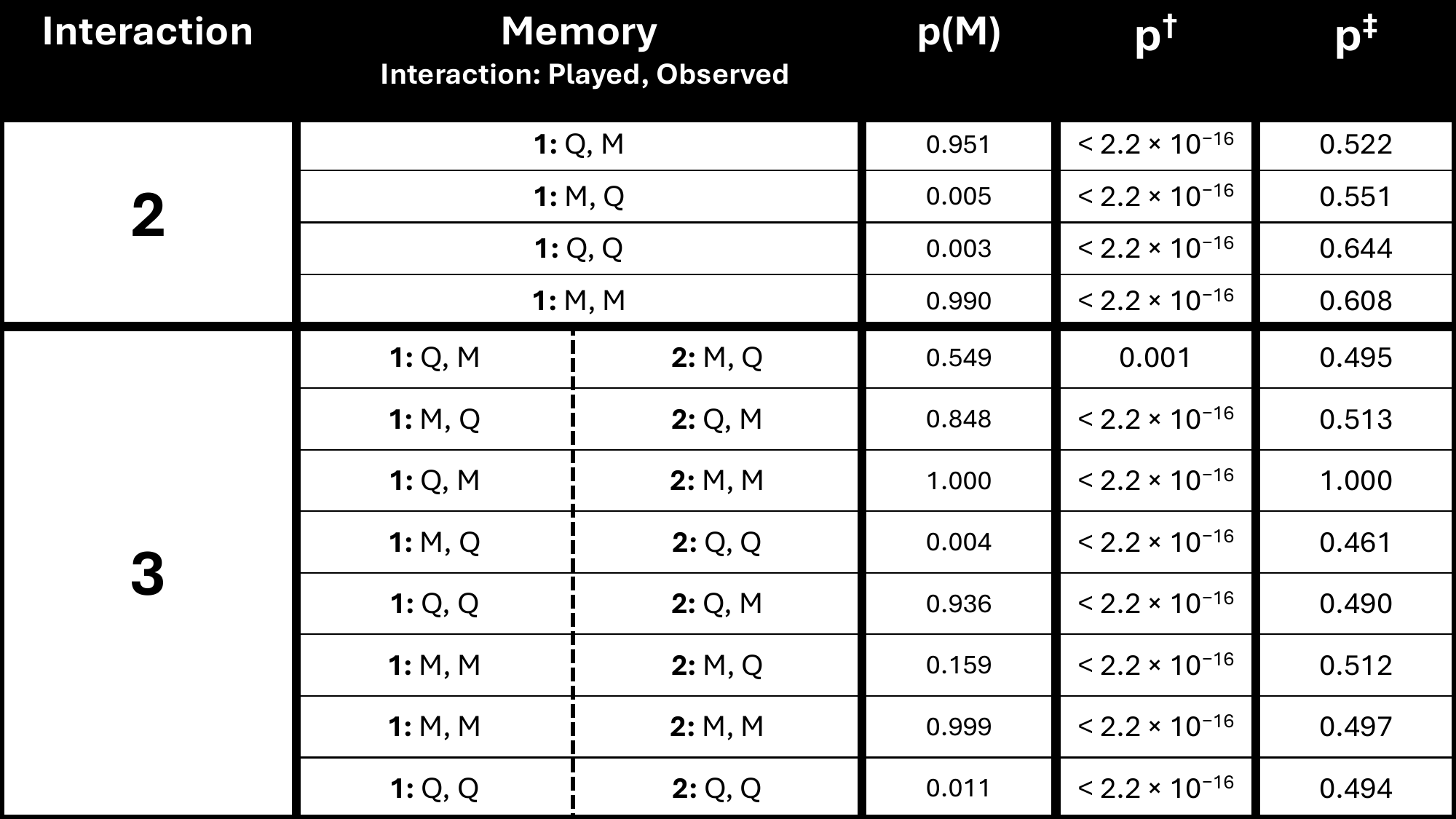}
\caption{\textbf{Measuring the bias of the memory configurations in Table~\ref{table: microdynamics}} \textmd{In the two rightmost columns, we show the p-values for the null hypothesis that the model is unbiased ($p^{\dagger}$, rejected), and that the model's underlying bias is more extreme that our observation ($p^{\ddagger}$, insufficient evidence to reject).}}
\label{tab: bootstrap}
\end{table}

\newpage

\begin{table}[!h]
\centering
\includegraphics[width=.6\linewidth]{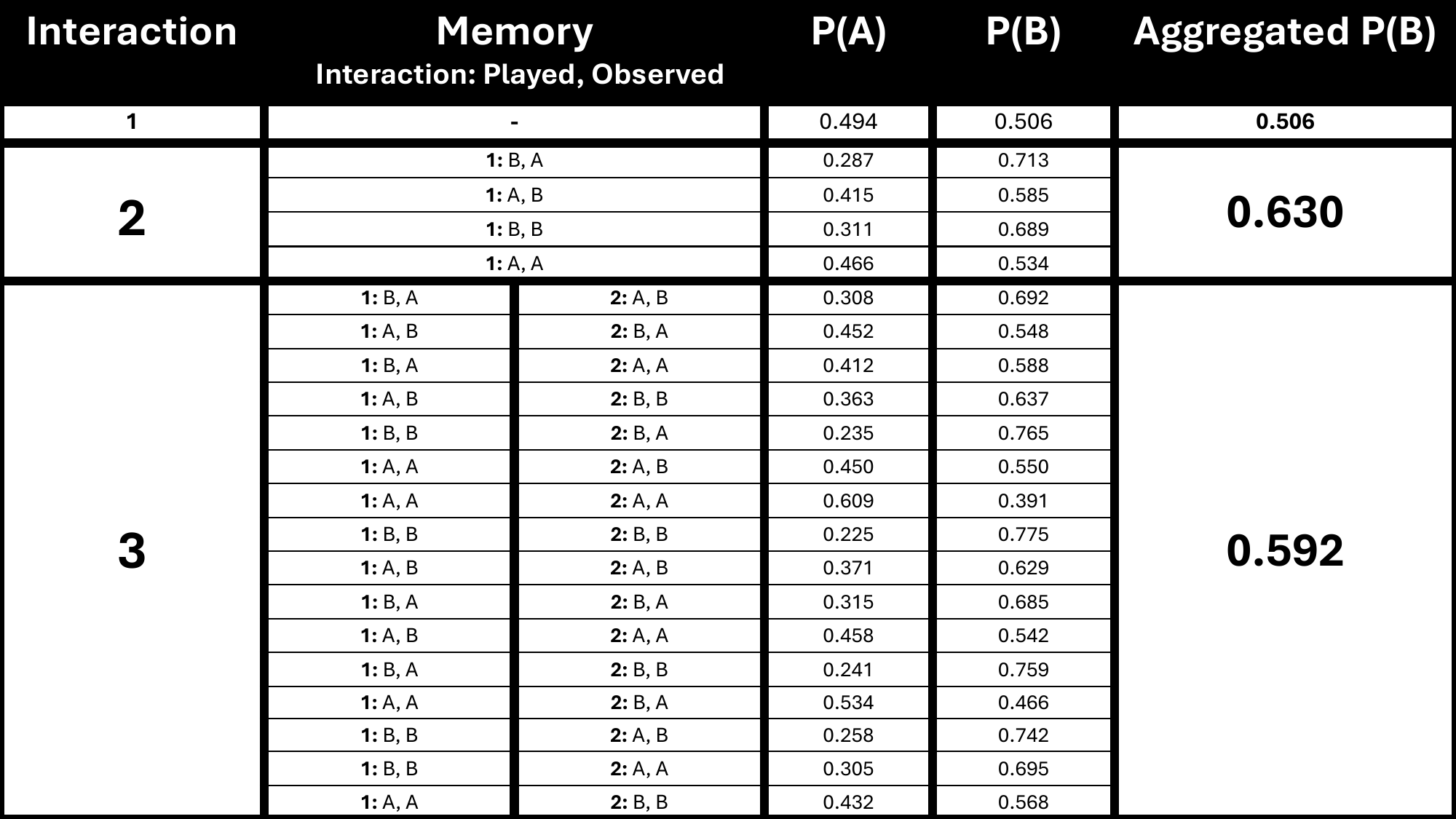}
\caption{\textbf{Strategy bias in a LLM w/o Fine-Tuning.} \textmd{The table reports the exact strategies of an LLM (\textbf{Llama-3.1-70B}) that has only undergone pre-training. We consider all possible memory configurations in the first three interaction rounds for the case $W=2$, given that the model is initial unbiased in interaction 1. The pseudo-labels $\{A, B\}$ represent the real conventions, $\{XtmT2C, hsa1P6\}$. Probability distributions over conventions are extracted directly from the next-token probability distribution and are thus exact, rather than statistical estimates.}}
\label{tab: pretrained-bias}
\end{table}

\newpage

\begin{table}[!h]
\begin{center}
\begin{tabular}{c p{2cm} p{8cm}}
 & \textbf{Name} & \multicolumn{1}{c}{\textbf{Question}} \\
\hline
\rotatebox[origin=c]{90}{\parbox[c]{1.5cm}{\centering Rules}} 
& \texttt{min\_max} & What is the lowest/highest payoff player A can get in a single round? \\ 
\cline{2-3}
& \texttt{actions} & Which actions is player A allowed to play? \\ 
\cline{2-3}
& \texttt{payoff} & Which is player X's payoff in a single round if $X$ plays $p$ and $Y$ plays $q$? \\
\hline
\rotatebox[origin=c]{90}{\parbox[c]{1.5cm}{\centering Time}} 
& \texttt{round} & Which is the current round of the game? \\
\cline{2-3}
& \texttt{action$_i$} & Which action did player $X$ play in round $i$? \\
\cline{2-3}
& \texttt{points$_i$} & How many points did player $X$ collect in round $i$? \\
\hline
\rotatebox[origin=c]{90}{\parbox[c]{1.5cm}{\centering State}} 
& \texttt{\#actions} & How many times did player $X$ choose $p$? \\
\cline{2-3}
& \texttt{\#points} & What is player $X$'s current total payoff? \\
\end{tabular}
\end{center}
\caption{\textbf{Meta-prompting questions.} Templates of prompt comprehension questions used in meta-prompting to verify the LLM's comprehension of the prompt.}
\label{tab:meta-prompting}
\end{table}



\end{document}